\documentclass[pre,aps,amsfonts,amssymb,multicol,epsfig,graphics,referee]{revtex4}
\usepackage{graphicx}
\usepackage{bm}

\def \beq{\begin{equation}}         \def \eeq{\end{equation}}
\def \beqa{\begin{eqnarray}}        \def \eeqa{\end{eqnarray}}
\def \bea{\begin{array}}        \def \eea{\end{array}}

\def\bio#1#2#3{{Biophys. J. }{\bf #1}, #2 (#3)}

\def\jpa#1#2#3{{J. Phys. A: Math. Gen.}{\bf #1}, #2 (#3)}
 
\def\jcp#1#2#3{{J. Chem. Phys. }{\bf #1}, #2 (#3)}

\def\mol#1#2#3{{Macromolecules }{\bf #1}, #2 (#3)}
\def\nat#1#2#3{{Nature(London) }{\bf #1}, #2 (#3)}

\def\nats#1#2#3{{Nature struct. Biol. }{\bf #1}, #2 (#3)}
\def\pnas#1#2#3{{Proc. Natl. Acad. Sci. USA }{\bf #1}, #2 (#3)}
\def\pre#1#2#3{{Phys. Rev. E }{\bf #1}, #2 (#3)}
\def\pra#1#2#3{{Phys. Rev. A }{\bf #1}, #2 (#3)}

\def\prl#1#2#3{{Phys. Rev. Lett. }{\bf #1}, #2 (#3)}
\def\sci#1#2#3{{Science }{\bf #1}, #2 (#3)}

\usepackage{amsmath}
\usepackage{epsfig}

\begin{document}

\title{A constant extension ensembles model of double-stranded chain molecules} 
\author{Fei Liu}
\email[]{liufei@itp.ac.cn}
\affiliation{Institute of Theoretical Physics, The Chinese 
Academy of Sciences, P. O. Box 2735, Beijing 100080, China}
\author{Luru Dai}
\affiliation{Institute of Theoretical Physics, The Chinese 
Academy of Sciences, P. O. Box 2735, Beijing 100080, China}
\author{Tao Xu}
\affiliation{Institute of Theoretical Physics, The Chinese 
Academy of Sciences, P. O. Box 2735, Beijing 100080, China}
\author{Zhong-can Ou-Yang}
\affiliation{Institute of Theoretical Physics, The Chinese 
Academy of Sciences, P. O. Box 2735, Beijing 100080, China}
\affiliation{Center for Advanced Study, Tsinghua University, Beijing 100084, 
China}

\date{\today}
\begin{abstract}
Because the constant extension ensemble of single chain molecule is  
not always equivalent with constant force ensemble, a model of 
double-stranded conformations, as in RNA molecules and $\beta$-sheets in 
proteins, with fixed extension constraint is built in this paper. Based 
on polymer-graph theory 
and the self-avoiding walks, sequence dependence and excluded-volume 
interactions are explicitly taken into account. Using the 
model, we investigate force-extension curves, contact distributions and 
force-temperature curves at given extensions. We find that, for the same 
homogeneous chains, the 
force-extension curves are almost consistent with the extension-force curves 
in the conjugated force ensembles. Especially, the consistence depends on 
chain lengths. But the curves of the two ensembles are completely 
different from each other if sequences are considered. In addition, contact 
distributions of homogeneous sequence show that the double-stranded regions in 
hairpin conformations tend to locate at two sides of the chain. We 
contribute the unexpected phenomena to the nonuniformity of excluded-volume 
interactions of the region and two tails with different lengths. This tendency will 
disappear if the interactions are canceled. Finally, in constant 
extension ensemble, the force-flipping transitions conjugated with re-entering 
phenomena in constant force ensemble are observed in 
hairpin conformations, while they do not present in secondary structure 
conformations. 
\end{abstract}
\maketitle

\section{introduction}
\label{sec-intro}
In recent years, advances in manipulation techniques have made it possible to 
measure and characterize biological macromolecules at single-molecule level. By 
using devices such as optical and magnetic tweezers, atomic force 
microscopy, basic mechanical, physical and chemical 
properties of fundamental biological objects, e.g., proteins, nucleic 
acids, and molecular motors were obtained\cite{smith96,maier,zlatanova, 
riefp, visscher}. 
Special efforts have been devoted to mechanical properties of the nucleic 
acids, from elastic stretching experiments early\cite{smith96} to 
recently unzipping double-stranded DNA (dsDNA), single-stranded DNA (ssDNA) 
or RNA\cite{maier,bockelmann,liphardt}. Many useful insights can be 
obtained by analyzing the extension-force curves (EFCs) or 
force-extension curves (FECs) recorded in experiments, e.g., S-dsDNA 
structure found\cite{smith96,rief} and the measurement for the proportion 
of G/C compared to A/T contents along dsDNA sequence\cite{bockelmann}.

On the theoretical side, a number of models have been built to interpret or 
simulate nucleic acid mechanical experimental data and phenomena. Elastic 
properties of stretched dsDNA were described by Marko and Siggia\cite{marko}, 
and Zhou {\it{et al.}}\cite{zhou1}. Montanari and 
M$\acute{e}$zard revealed that a second order phase transition exhibits as 
stretching ssDNA\cite{montanari}. Lubensky and Nelson presented an 
extensive theoretical investigation of mechanical unzipping 
dsDNA\cite{lubensky}. Gerland {\it{et al.}} explored quantitatively how 
secondary structures determine outcome of FECs\cite{gerland}. Although great 
efforts have been contributed to DNA mechanical problem, we 
note that little theoretical works concerned about constant extension 
ensembles of ssDNA or RNA\cite{gerland}, in which two ends separation 
$\bf{r}$ of chain is fixed, and the average force is 
measured\cite{bockelmann}. It is known for a time that in a traditional single 
polymer system, the constant force and the constant extension ensembles are 
equivalent in thermodynamic limit, though for finite systems inequivalence 
might be expected\cite{titantah}. For nucleic acid, however, the 
situation becomes more complex since the presence of monomer interactions 
and the absence of self-averaging arising from sequences\cite{lubensky,bhat}. 
To study such ensembles further, we will modify and extend the statistical 
model of double-stranded chain conformations developed by Chen and 
Dill\cite{chen95,chen98} to fixed extension scenario. To be different from 
previous theories\cite{gerland}, the extended model retains a relatively 
high degree of realism, which the sequence dependence and excluded-volume (EV) 
interactions are took into account explicitly. Experiments have shown that 
sequences change EFCs or FECs dramatically\cite{maier,bockelmann,liphardt,
rief}; while very recent stretching ssDNA experiment revealed the importance 
of EV interactions\cite{dessinges}. The interactions were neglected 
before. On the other hand, the model is more ``microscopic", i.e., entropy 
of chain is computed from the number of conformations directly. As the first 
step, we restrict the model on two dimension (2d) lattice; and the extension 
is specified to be one component of the separation $\bf{r}$ for simplicity. 
More general constant $\bf{r}$ will be given in future work. 

The organization of the paper is as follows. In Sec.~\ref{reviewpolymer}, 
the statistical model of double-stranded chain molecules is simply 
reviewed. In Sec.~\ref{extensiondb}, we extend two classes of 
conformations, hairpin and RNA-like secondary structure to constant 
extension cases, respectively. In Sec.~\ref{discussion}, we first 
investigate how FECs of two classes of conformations are changed with 
temperature, chain length and sequence. To relate FECs with molecule 
structures, monomer-monomer contact probability distributions are then 
introduced. As an illustration, the distributions of simpler hairpin 
conformations are calculated. It is unexpected to find that hairpin contacts 
tend to form at two sides of chain. To explore the underlying reason, 
we introduce the Asymmetric Function (AF), and find that the interesting 
phenomena are the results of nonuniform EV interactions of double-stranded 
regions and two tails with different lengths in hairpin conformations. At last part 
of this section, force-temperature curves (FTCs) at fixed extensions for 
homogeneous double-stranded chains are computed. This study comes from 
re-entering transitions observed in constant force ensembles\cite{marenduzzo,
orlandini}. Force flipping phenomena are observed in hairpin conformations, 
but do not present in secondary structure. The ``flipping" here is defined 
that when temperature decreases, force value first decreases and then 
increases; it finally decreases again as temperature is lower than some value. 
We believe that the phenomena in constant extension ensembles are conjugated 
with re-entering transition. Section~\ref{conclusion} is our conclusion. 
The calculation of extensions with a special case in hairpin conformations is 
relegated in Appendix~\ref{endtoendhairpin}.  

\section{The double-stranded chain model}
\label{reviewpolymer}
The details of the statistical model of double-stranded chain molecules 
can be found in Refs.~\cite{chen95} and \cite{chen98}. Here we just give 
a brief review.

\subsection{Polymer graph theory}
\begin{figure}[htpb]
\begin{center}
\includegraphics[width=0.5\columnwidth]{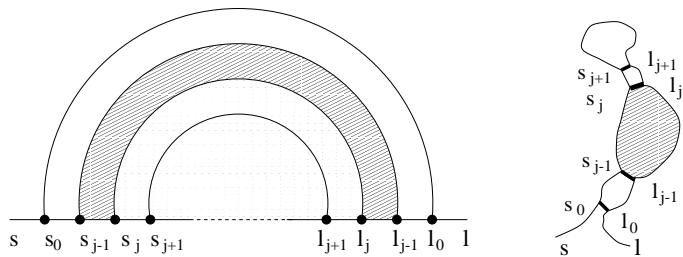}
\caption{The polymer graph of hairpin conformations. The shadowed region is 
a face of the graph. The graph is divided into two parts: two tails 
$[s,s_0]$ and $[l_0,l]$, and one double-stranded region, or CG  
enclosed by outmost link $(s_0,l_0)$. Here $s$ and $l$ are two end monomers in 
chain. The double-stranded region is composed of links nested each other: 
$(s_0,l_0),\cdots (s_{j-1},l_{j-1}),(s_j,l_j), (s_{j+1},l_{j+1})$.}
\label{polygraph}
\end{center}
\end{figure}

The model is based on polymer 
graphs, diagrammatic representations of the self-contacts made by 
different chain conformations. Fig.~\ref{polygraph} shows a hairpin 
conformation and corresponding polymer graph: vertices 
represent the chain monomers, straight line links symbolize the covalent 
bonds, and curved links stand for spatial contact between monomers.
A given polymer graph represents an ensemble of chain conformations that are 
consistent with given contacts. Conformations having contacts other 
than those specified by the polymer graph belong to other graphs. 
In general, fewer curved links exist in a graph, a more larger number of 
chain conformations are consist with this graph. Any two pairs of curved 
links in a polymer graph must bear one of three relationships: nested, 
unrelated and crossing linked. 
Graphs of the double-stranded chain conformations involve no crossing 
links, examples of which include the simplest hairpin structure, such 
as dsDNA, and mainly secondary structures among nucleic acids and antiparallel 
$\beta$-sheets in proteins. In terms of polymer graph, the partition function 
of a $(N+1)$ monomers ((N+1)-mer) chain molecule is given as a sum over all 
possible polymer graphs, 
\begin{eqnarray}
\label{graphsum}
{\cal Q}_N(T)=\sum\limits_{\Gamma}\Omega(\Gamma)e^{-E(\Gamma)/k_B T},
\end{eqnarray}
where $k_B$ is Boltzmann constant, $T$ is temperature, $\Gamma$ is an index of a possible polymer graph, $E(\Gamma)$ and $\Omega(\Gamma)$ are the energy 
and the number of conformations  of the given polymer graph $\Gamma$, 
respectively. To calculate $\Omega(\Gamma)$, 
Chen and Dill developed a matrix multiplication method\cite{chen95,chen98}.

\subsection{The matrix method for a given polymer graph}
\begin{figure}[htpb]
\begin{center}
\includegraphics[width=0.7\columnwidth]{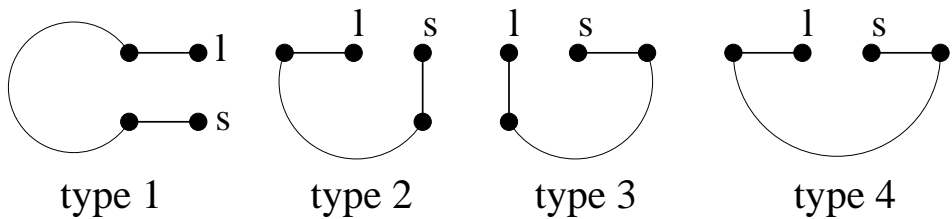}
\caption{ The four types of port (inlet or outlet) shapes on 2D lattice.} 
\label{portshap}
\end{center}
\end{figure}

A complex polymer graph can be divided into a 
series of faces consecutively, in which  
face is a region in graph that is bounded by curved and straight lines 
and contains no other edges; see the shadowed region in Fig.~\ref{polygraph}. 
Faces are classified into five types: left (L), middle (M), right (R), 
left-right (LR) and isolated (I), according to the arrangement of the nested 
curved links that bound the face. The calculation of the full partition 
function $\Omega(\Gamma)$ for a given  graph $\Gamma$ is correspondingly 
separated into two steps: to count all conformations for each face, as if 
they are isolated and independent of each other, and to assemble these  
conformations into $\Omega(\Gamma)$. To avoid that conformations of different 
faces bump into each other (EV interaction), more detailed 
information about the conformations of faces is needed. On two dimension (2D) 
lattice, it is realized by exact enumeration. First, conformations of each 
face are classified into sixteen types according to the shape of ports 
(inlet and outlet) through which it is connected to other faces. The shapes 
of ports are shown in Fig.~\ref{portshap}. Then the compatibility between 
neighboring faces 
can be checked exactly through the spatial compatibility between the outlet 
of one face and the inlet of next face. It is convenient to 
introduce two matrices: the Face Count Matrix (FCM) ${\bf S}_t$, 
which matrix element $({\bf S}_t)_{ij}$ is the number of conformations having 
an inlet conformation of type $i$ ($1\le i\le 4$) and an outlet 
conformation of type $j$ ($1\le j\le 4$) for a type $t$ face, and 
viability matrix ${\bf Y}_{t_1t_2}$, which element 
$({{\bf Y}_{t_1t_2}})_{ij}$ is $1$ or $0$ if connection 
of type $i$ outlet of a type $t_1$ face and type $j$ inlet of a type 
$t_2$ face is viable or not viable.  For a hairpin graph $\Gamma$ having 
$M$ faces, $\Omega(\Gamma)$ is derived as a product of matrices:  
\begin{eqnarray}
\label{omegcalculate}
\Omega(\Gamma)={\bf U}\cdot{\bf S}_{t_M}\cdot{\bf Y_{t_Mt_{M-1}}}\cdot{\bf S}_{t_{M-1}}\cdots{{\bf S}_{t_1}}\cdot{\bf U}^t,
\end{eqnarray}
where ${\bf U}=\{1,1,1,1\}$, ${\bf U}^{t}$ is the transpose of 
${\bf U}$. 

\subsection{The partition function}
In order to calculate the partition function of a whole chain using 
Eq.~\ref{graphsum}, the sum over all possible polymer graphs is 
necessary. For the double-stranded conformations, an efficient dynamic 
programming algorithm was developed in Ref. ~\cite{chen98}. 
The idea is to start with a short chain segment and elongate the segment by 
adding one monomer for each step, and to calculate recursively the partition 
function of the longer segment using the result of the shorter one. 
The algorithms will be discussed as needed in following sections. More 
useful alternative expression of Eq.~\ref{graphsum} is 
\begin{eqnarray}
\label{engsum}
{\cal Q}_N(T)=\sum\limits_E g_N(E)e^{-E/k_B T},
\end{eqnarray}
where $g_N(E)$ is the density of states, or the number of conformations having  
energy $E$, which is defined as 
\begin{eqnarray}
g_N(E)=\sum\limits_{\Gamma}\left.\Omega(\Gamma)\right|_{E(\Gamma)=E}.
\end{eqnarray}

\section{Constant extension ensembles for double-stranded chain molecules:  
hairpin and secondary structure conformations}
\label{extensiondb}

\begin{figure}[htpb]
\begin{center}
\includegraphics[width=0.8\columnwidth]{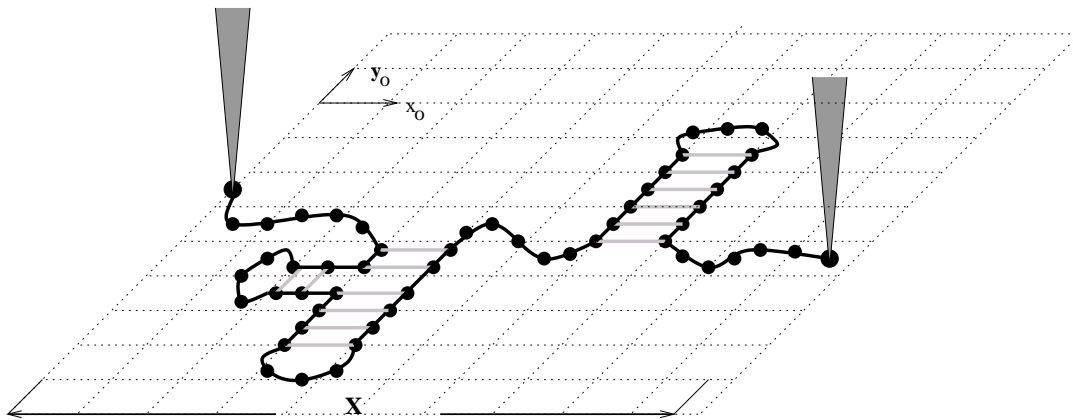}
\caption{Sketch of the constant extension experiment on the 2D lattice. The 
larger dark points represent two ends of a chain molecule; they are grasped 
by two pins. Monomers are denoted to be small dark points. In present paper, 
only projection component $x$ of separation $\bf{r}$ along direction 
${\bf x}_o$ is fixed as constant. Average force $f$ along $\bf{x}_o$ is 
recorded as a function of $x$.}
\label{sketch}
\end{center}
\end{figure}

Fig.~\ref{sketch} depicts the situation studied in this paper. Two ends of 
chain molecule are grasped by two pins. Instead of fixing end-to-end 
distance (EED)$\bf{r}$, its projection along direction ${\bf x}_o$, or 
extension $x$ is required to be constant. Average force $f$ along ${\bf x}_o$ 
is recorded as a function of $x$. Because of extension constraint, 
the partition function Eq.~\ref{engsum} is modified to ${\cal Q}_N(x;T)$ 
\begin{eqnarray}
\label{engsumforc}
{\cal Q}_N(T;x)=\sum\limits_E g_N(E;x)e^{-\beta E},
\end{eqnarray}
where $g_N(E;x)$ is the number of conformations whose extensions are 
$x$. Then force $f$ can be calculated by   
\begin{eqnarray}
\label{flory}
f(x,T)=-k_BT\frac{\partial}{\partial x}\log{\cal Q}_N(x;T). 
\end{eqnarray}
Considering that our chain model is on 2D lattice, in following sections 
constant extension $x$ is to be replaced by a discrete variable $\Delta$. 

The appearance of parameter, extension $\Delta$ requires the statistical model 
of double-stranded chains to be modified and extended carefully. In the next 
two sections, we will show how to compute $g_N(E;\Delta)$ of hairpin and 
secondary structure conformations, respectively. 

\subsection{Constant extension ensembles of hairpin conformations}
\label{fixhp}
As one of the simplest elements in secondary structure, hairpin 
conformations  
exist in a large class of biomolecules, such as RNA hairpin, peptide 
$\beta$-hairpin and DNA hairpin. Recent works showed that the 
hairpin conformations have remarkable thermodynamic and kinetic 
behaviors\cite{munoz,chen98}. In constant force ensembles, the mechanical 
behaviors of the hairpin conformations are completely different from that 
of secondary structure conformations\cite{liuf}. It is interesting to see 
whether there is new difference presented in constant extension ensembles. 
In addition, 
more precise formula and the essence of theory for complex secondary 
structure conformations make us to explore their properties independently. 

The polymer graphs of hairpin conformations are that every curved 
link bears a nested relationship with respect to every other curved link.
The polymer graph is composed of two parts: two non-self-contacting 
tail chains $(s,s_0)$ and $(l_0,l)$ and one double-stranded region 
$(s_0,l_0)$, which is defined as Closed Graph (CG)\cite{chen98}; 
see Fig.~\ref{polygraph}. The number of conformations for a given graph 
equals a multiplication of the number of double-stranded 
conformations and the number of two tails conformations. In terms of 
four types of the outermost faces, the graphs are classified into four 
types. (LR type is excluded from hairpin conformations). To sum over all 
polymer graphs, two matrices, the Closed Graph Count Matrix 
(CGCM), ${{\bf G^\ast}_t}\left[E,s_0,l_0\right]$ and diagonal matrix 
${\bf \omega}\left[s,s_0;l_0,l\right]$ have been defined: 
$({{\bf G^\ast}_t}\left[E,s_0,l_0\right])_{ij}$ is the sum over the number 
of conformations for all possible $t$ type graphs having energy $E$, 
given that the outmost link spans from vertex $l_0$ to vertex $s_0$ and 
innermost and outermost links are in $i$ and $j$ types conformations; 
diagonal matrix element $({\bf \omega}\left[s,s_0;l_0,l\right])_{ii}$ is the 
number of conformations of two tails $(s,s_0)$ and $(l_0,l)$ that are 
spatially compatible with type $i$ conformations. State density $g_N(E)$ then 
can be written as: 
\begin{eqnarray}
g_N(E)={\bf U}\cdot\sum_{s_0, l_0, t}{\bf\omega}\left[s,s_0;l_0,l\right]
\cdot{\bf G^\ast}_t\left[E,s_0,l_0\right] \cdot{\bf U}^t,
\label{hairpinsum}
\end{eqnarray}
here $1\le t\le 4$\cite{chen98}. 

\begin{figure}[htpb]
\begin{center}
\includegraphics[width=1.\columnwidth]{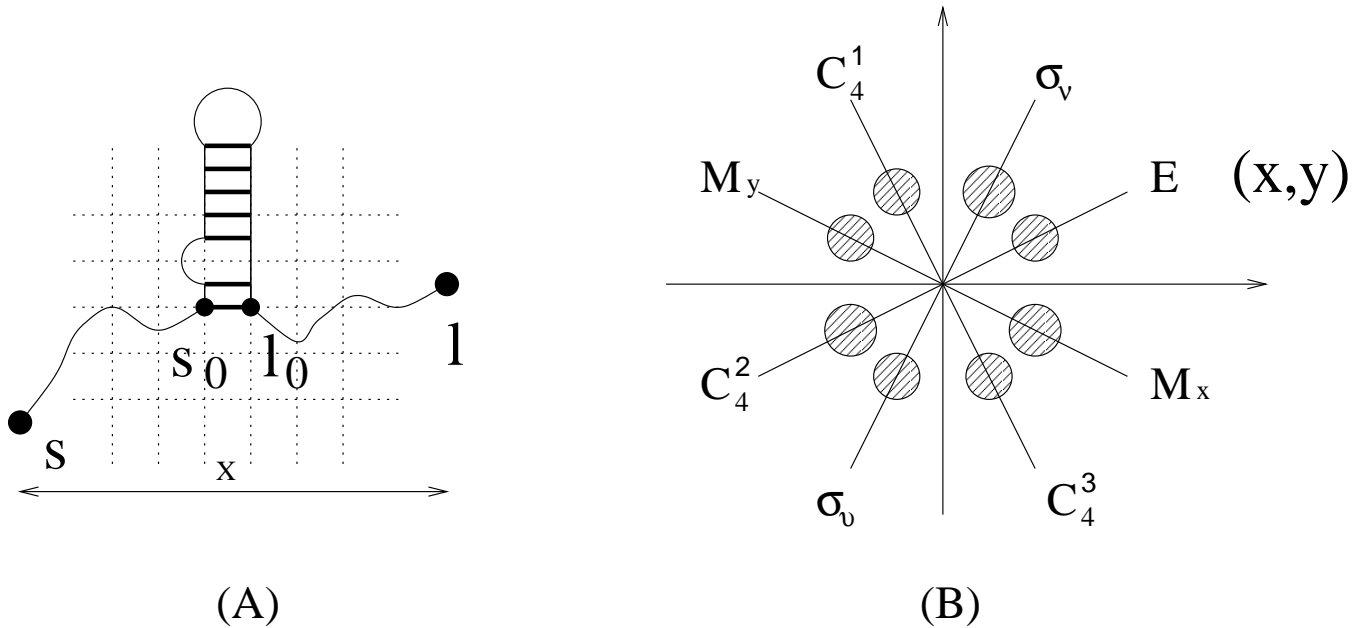}
\caption{(A) Sketch of hairpin conformations at constant extension $x$. 
Here the double-stranded region contributes $+1$ to extension of whole chain. 
(B) Illustration of how the conformation ended at $(x,y)$ is  
distributed to whole lattice plane by eight square symmetric transformations, 
where the shadowed circles represent the double-stranded regions. 
}
\label{hairpinextn}
\end{center}
\end{figure}

When the extension of hairpin conformations is fixed, $g_N(E)$ is then 
extended to $g_N(E,\Delta)$, and ${\bf \omega}[s,s_0;l_0,l]$ is 
{\it{nondegenerated}} into 
${\bf \omega}[s,s_0;l_0,l|\Delta]$. We rewrite Eq.~\ref{hairpinsum} as 
\begin{eqnarray}
g_N(E;\Delta)=
{\bf U}\cdot\sum_{s_0, l_0, t}{\bf\omega}\left[s,s_0;l_0,l|\Delta\right]
\cdot{\bf G^\ast}_t\left[E,s_0,l_0\right] \cdot{\bf U}^t.
\label{hairpinsum_1}
\end{eqnarray}
Fig.~\ref{hairpinextn}(a) shows that extensions of chain on 2D lattice do not 
involve any detailed CG structure directly, i.e., only the outmost link 
$(s_0,l_0)$ contributes $\pm 1$ or $0$ to $\Delta$. Though, in real 
nucleic acid, a typical distance of a hydrogen bond is about $3$ times 
than the nucleotide distance, it does not make sense for our description on 
a coarse-grained level. $\Delta$ is enumerated with three steps: 
first, to fix one conformation of a given CG on 
lattice and grow two tails which are compatible with the graph type; then, to 
enumerate two tails extensions and combine them with distance of the graph 
according ${\bf x}_o$ direction; finally, to distribute the conformation 
to full lattice plane by eight square symmetry transformations ($D_4$ group); see 
Fig.~\ref{hairpinextn}(b). The basic idea is very 
simple, however, it is very cumbersome to complete this process. Rather than 
to give a detailed analysis here, the process is illustrated in 
Appendix~\ref{endtoendhairpin} by the simplest case. 

Although the enumeration method gives quite accurate hairpin extensions, 
unfortunately, it is impossible to count longer chain. We propose a 
more practical approximation in the next section. 

\subsection{Constant extension ensembles of secondary structure 
conformations}
Secondary structure can be seen as a tree-like structure into which four 
basic structure elements, helix, loops, bulges and junctions compose 
through self-similarity arrangement.  Each graph of secondary structure 
conformations is divided into two levels: the first lever is a combination 
of unrelated double-stranded regions connected with single-stranded chains; 
the second level is that each CG may be viewed as an independent 
secondary structure except that two end monomers of the region contact.  
We first simply review how to calculate the state density $g(E)$ without 
mechanical constraint. The necessary definitions are introduced.  

To be different from hairpin conformations, the calculation of full state 
density of the secondary structure is more complex. First all possible CGCMs, 
${{\bf G^\ast}_t}\left[E,a,b\right]$ (Note that LR-type faces are included) 
are computed, where $(a,b)$ is the outmost link of 
the CG. Any CG is composed of smaller 
unrelated subclosed graphs, auxiliary matrices, ${\bf K}_t[E,l,a,b]$ counting 
a combination of conformations of all subgraphs are introduced, where 
${\it cycle length }$ $l$ is the total number of monomers of the 
single-stranded chains in $[a,b]$\cite{chen98}. CGCMs can be obtained by 
multiplication of matrix ${\bf K}_t[E,l,a,b]$ and the number of 
conformations of the single-stranded part. Then in order to combine 
conformations of unrelated graphs into whole, 
matrices ${\bf G}_t\left[E,s,a\right]$ having energy $E$ for full polymer 
graph $[s,a]$ are defined, where $t=0,1,2$ represent three full graph 
types which are classified depending on whether $0$, $1$ and $2$ existing 
links are connected to the rightmost monomer, see Fig.~\ref{recurse}. 
Their element 
$({\bf G}_t\left[E,s,a\right])_{ij}$ is the number of conformations for 
graphs in which the outmost link (of rightmost subgraph) and the 
innermost link (of the leftmost subgraph) are in $j$ and $i$ conformations, 
respectively. All matrices mentioned above are calculated by 
dynamics programming algorithm\cite{chen98}. The full density of states of 
the secondary structure conformations is written as
\cite{chen98} 
\begin{eqnarray}
g_N(E)={\bf U}\cdot\sum_{t=0}^{2}{\bf G}_t\left[E,s,l\right] \cdot{\bf U}
^t. 
\label{secstrusum}
\end{eqnarray}

\begin{figure}
\begin{center}
\includegraphics[width=0.6\columnwidth]{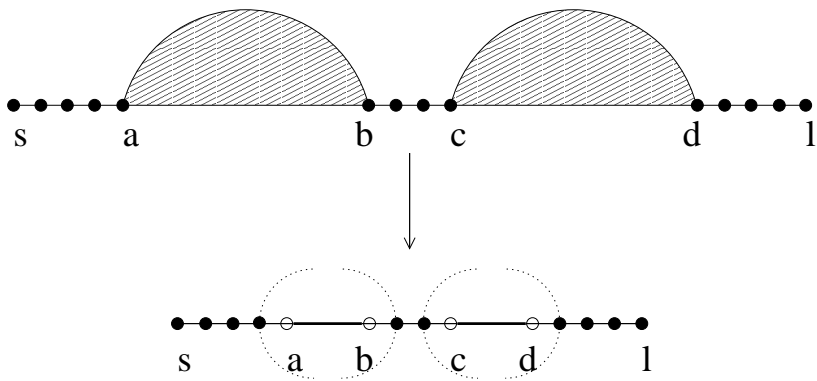}
\caption{ Illustration of how unrelated closed polymer graphs are reduced into 
effective chain to calculate the number of conformations with fixed extensions. 
For example, 
two closed graphs $[a,b]$ and $[c,d]$ in upper are replaced by two bonds 
in below, and 
length of the effective chain is reduced to $(a-s)+(c-b)+(l-d)+2$ from $l-s$. 
Considering that EV interactions of CGs and single-strand 
parts make conformations of the neighboring monomers of CGs to be ``frozen", 
it may be reasonable to reduce chain length more, showing here by dashed 
brackets. In present paper, the simplest case is considered. }
\label{renormchain}
\end{center}
\end{figure}

\begin{figure}[htpb]
\begin{center}
\includegraphics[width=0.8\columnwidth]{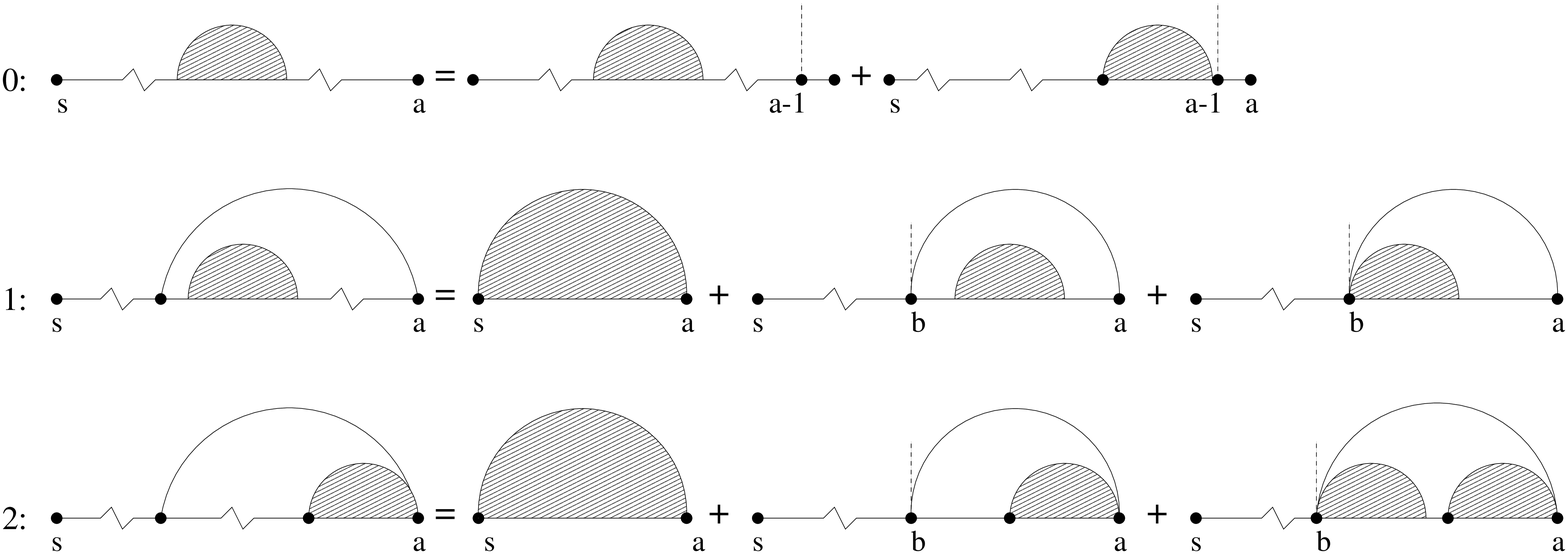}
\caption{the recursive relations for matrices ${\bf G}_t[E,n,s,a]$, 
where $t=0,1,2$.}
\label{recurse}
\end{center}
\end{figure}

Consider the number of conformations with fixed extension $\Delta$. 
Because that the detailed structures of CGs do not affect the whole 
extensions directly as in hairpin case, CGs are viewed as {\it effective} 
covalent bonds connecting left and right parts of a chain; the 
effective bonds bear the EV interaction of CGs with other 
units in the chain. Thus the secondary structure conformations are 
identified to be ``open" self-avoiding walks (OSAWs) with reduced monomers; 
see Fig.~\ref{renormchain}. OSAWs are self-avoid walks 
involving no neighboring contacts\cite{chen98,liuf}. The number of OSAWs  
whose extensions are $\Delta$ can be computed by enumeration and extrapolation 
method\cite{liuf}. 
Although the effective chain approach (ECA) may overestimate the number 
of conformations for partially counting EV interactions, 
we think that it is valuable before better methods are found.

To realize the ECA, we modify the matrices 
${\bf G}_t\left[E,s,a\right]$ to ${\bf G}_t\left[E,n,s,a\right]$, where 
new parameter $n$ is the number monomers of the effective chain. The recursive 
relations in Ref.~\cite{chen98} (also see Fig.~\ref{recurse}) are extended 
into following relations:
\begin{eqnarray}
{\bf G}_0[E,n,s,a]&=&{\bf G}_0[E,n-1,s,a-1]+{\bf G}_1[E,n-1,s,a-1],\\
{\bf G}_1[E,n,s,a]&=&
\delta_{n,1}\sum\limits_{t=L,M,I}{\bf G^\ast}_t[E,s,a]\nonumber\\ 
&+&\sum\limits_{0<b<a}\sum\limits_{E_1}{\bf G}_0[E-E_1,n-1,s,b]
\sum\limits_{t=L,M,I}{\bf G^\ast}_t[E_1,b,a]\nonumber\\
&+&\sum\limits_{0<b<a}\sum\limits_{E_1}{\bf G}_1[E-E_1,n-1,s,b]
\sum\limits_{t=M,I}{\bf G^\ast}_t[E_1,b,a],\\
{\bf G}_2[E,n,s,a]&=&
\delta_{n,1}\sum\limits_{t=R,LR}{\bf G^\ast}_t[E,s,a]\nonumber\\
&+&\sum\limits_{0<b<a}\sum\limits_{E1}{\bf G}_0[E-E_1,n-1,s,b]
\sum\limits_{t=R,LR}{\bf G^\ast}_t[E_1,b,a]\nonumber\\
&+&\sum\limits_{0<b<a}\sum\limits_{E_1}{\bf G}_1[E-E_1,n-1,s,b]
{\bf G^\ast}_{R}[E_1,b,a].
\end{eqnarray}
Correspondingly, $g_N(E)$ is extended to $g_N(E;\Delta)$ as 
\begin{eqnarray}
g_N(E;\Delta)=\sum\limits_{n=1}^{N} C_x(n;\Delta) 
\sum\limits_{t=0}^{2}{\bf U}\cdot{\bf G}_t[E,n,s,l]\cdot{\bf U^t}
\label{secstrusumnew}
\end{eqnarray}
where $C_x(n;\Delta)$ is the number of conformations of $n$-step OSAWs whose 
final $x$ coordinates are $\Delta$. Eq.~\ref{secstrusumnew} clearly 
separates contribution of the unrelated CGs from the single-stranded parts. 

Because the hairpin is one of four elements in secondary structure 
case, ECA should be suitable to it. We also note that exact enumeration 
still is available when the lengths of two tails are smaller than some 
value ($27$ in this paper), a mixture of two 
methods is applied in hairpin case. It is reasonable, for as the CG size is 
so large, or tails are so short that tail conformations are almost ``frozen", 
enumeration method is accurate. As the lengths of tails are relatively longer 
than CG size, or the EV interactions of CG and tail are smaller, the ECA may 
be available. 

\section{Tests of the model: FECs, contact distributions and 
force flipping phenomena}
\label{discussion} 
We have described how to extend the statistical mechanical model of 
double-stranded chain molecules to constant extension ensembles in previous 
sections. In this section, the model will be used to explore FECs, contact 
distribution, and force flipping phenomena. To account for specific monomer 
sequences, sequences are divided into two classes: homogeneous and RNA-like 
chain. Each contact of the homogeneous chain contributes one sticking 
energy $-\varepsilon$ ($\varepsilon<0$). While RNA-like chain has a 
specific sequence of four types of monomers: A, U, C and G, resembling the 
4 types of bases of an RNA; only A-U pair or C-G pair contributes one sticking 
energy $-\varepsilon$. In following, we take the lattice spacing $b$. 

But before beginning with our discussion , the even-odd oscillations in 
partition function ${\cal Q}_N(T;\Delta)$ with $\Delta$ have to be canceled,  
which are the results of 2D lattice restriction\cite{fisher66}. We damp the 
oscillations out simply by forming square root means of partition function as  
\begin{eqnarray}
\mathcal{Z}_N(T;\Delta)= \left[{\cal Q}_N(T;\Delta)*
{\cal Q}_N(T;\Delta+1)\right]^{1/2}. 
\end{eqnarray}
${\cal Q}_N(T;\Delta)$ in Eq.~\ref{flory} then is 
replaced by $\mathcal{Z}_N(T;\Delta)$.  

\subsection{Force-Extension curves}
\begin{figure}[htpb]
\begin{tabular}{cc}
\includegraphics[width=0.4\columnwidth]{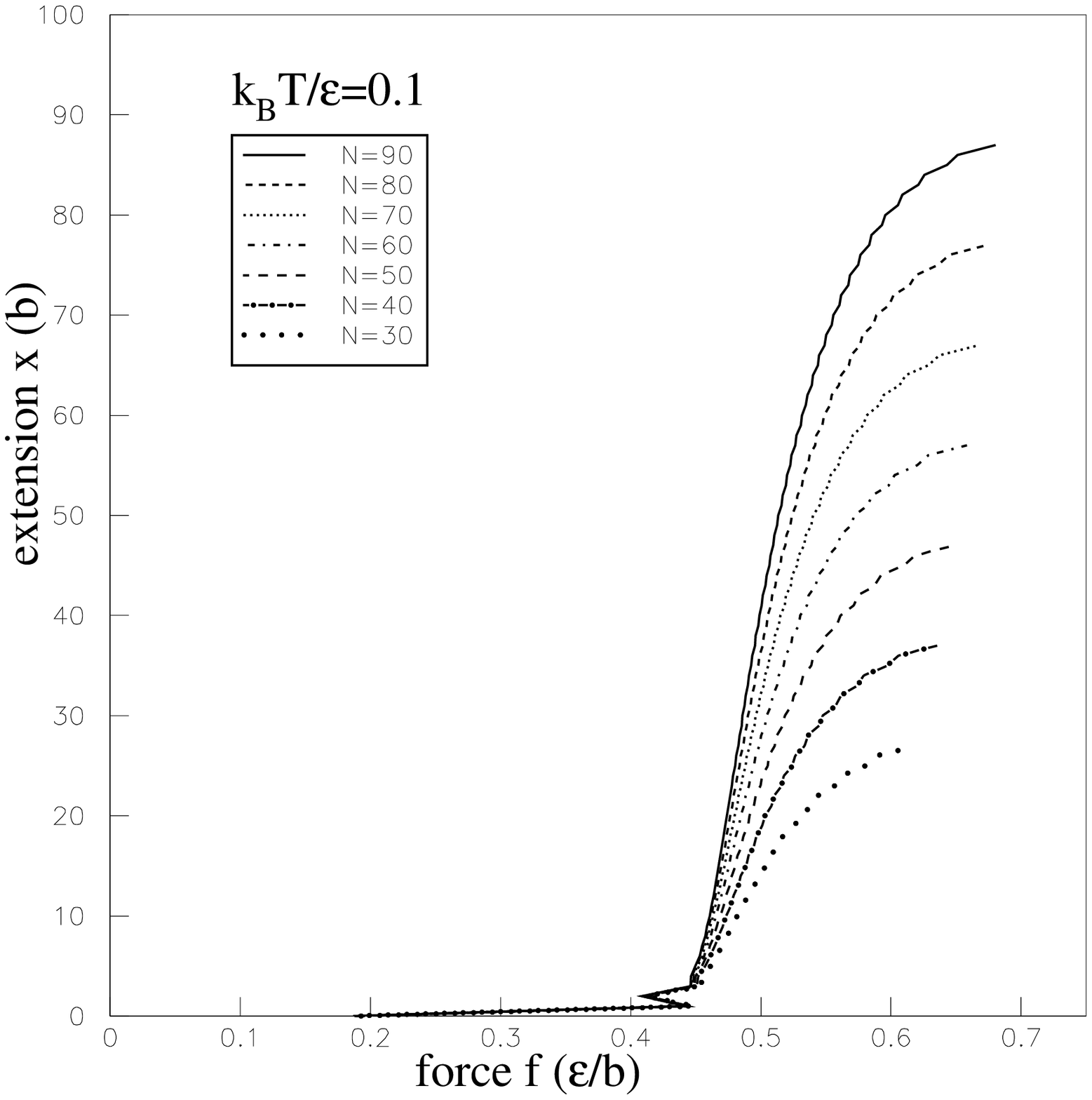}&
\includegraphics[width=0.4\columnwidth]{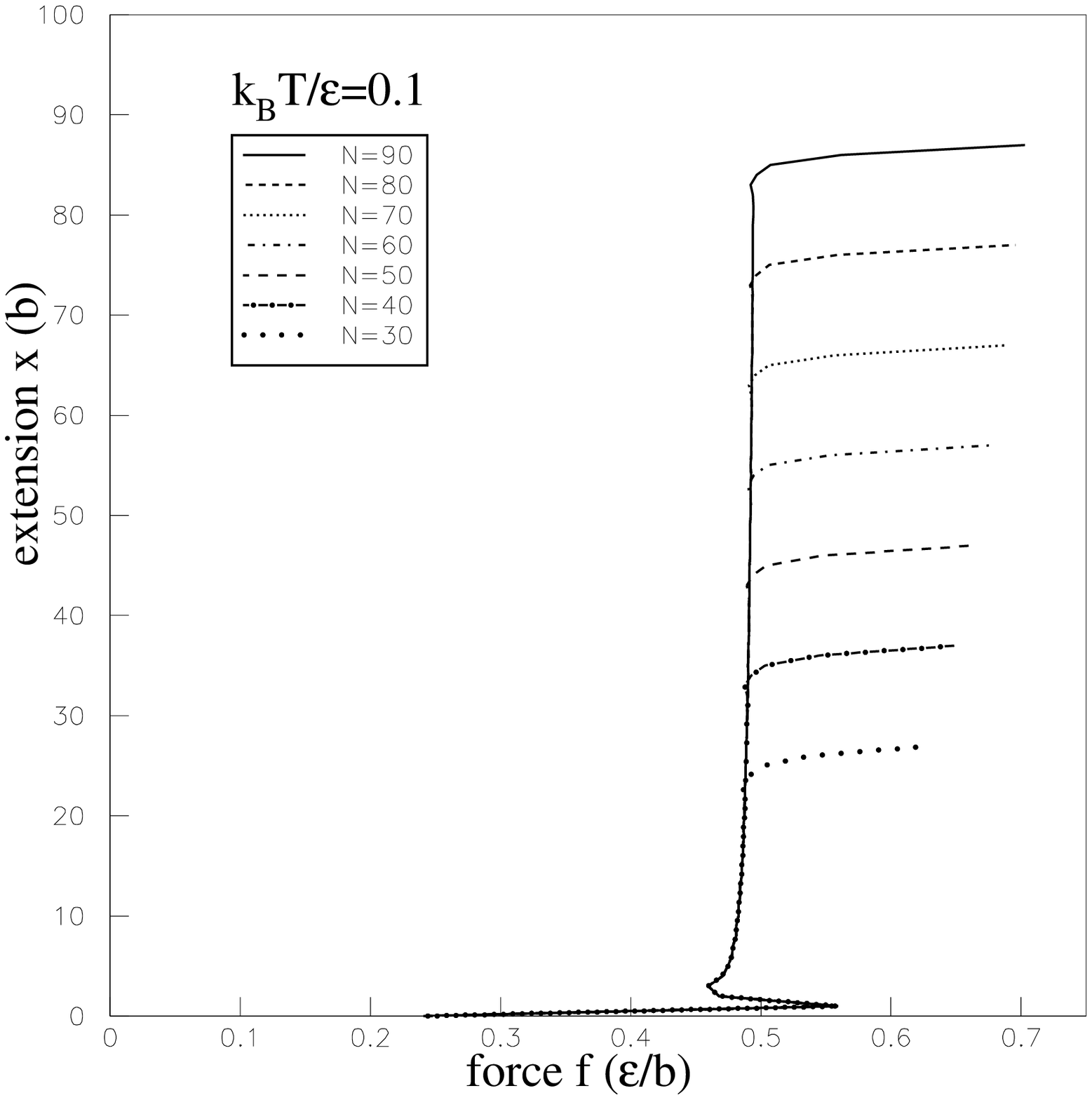}\\
(a)&(b)\\
\includegraphics[width=0.4\columnwidth]{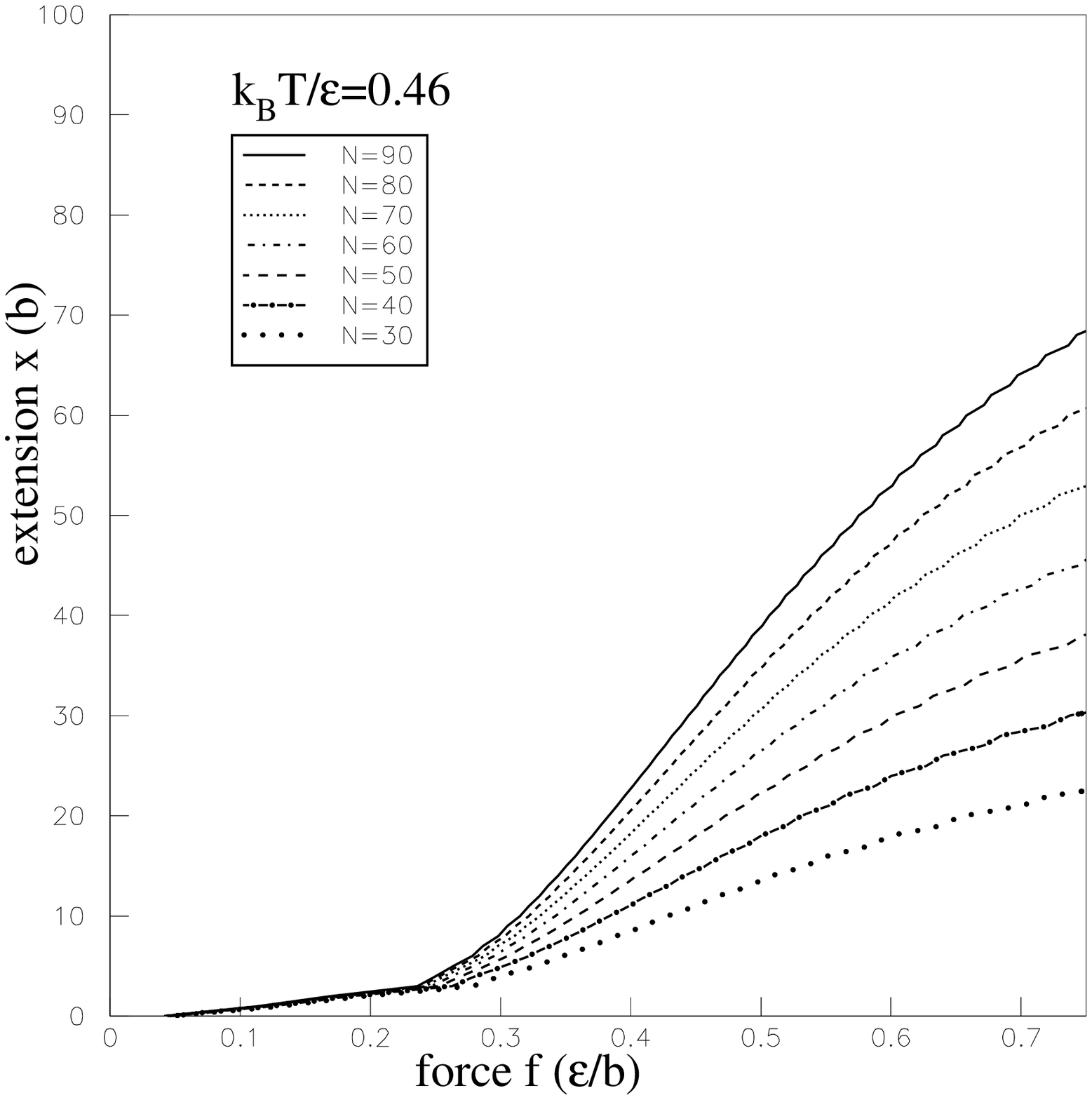}&
\includegraphics[width=0.4\columnwidth]{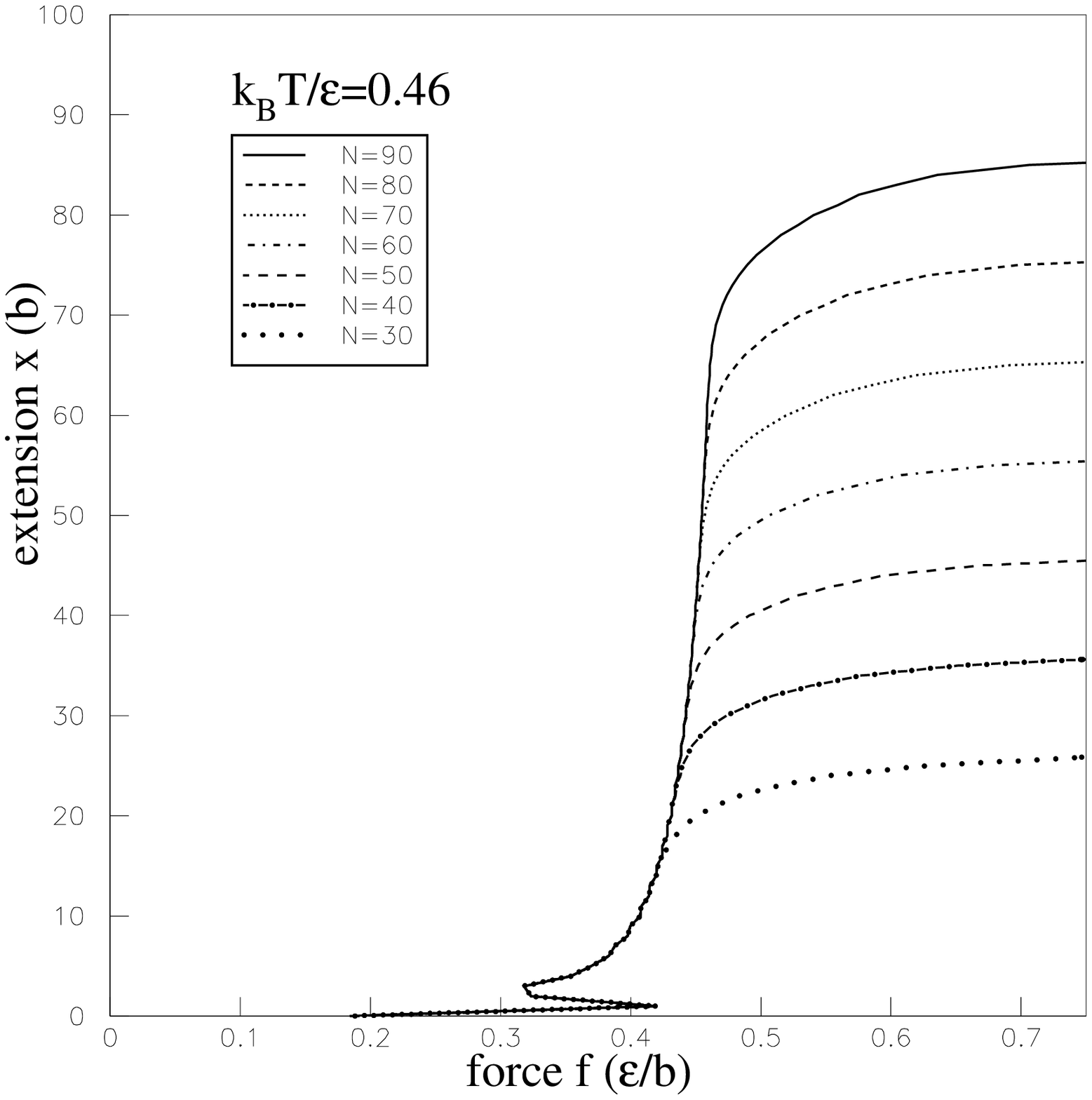}\\
(c)&(d)
\end{tabular}
\caption{The FECs of homogeneous chains restricted to secondary structure 
conformations: (a), (c), and hairpin conformations: (b), (d). To compare with 
EFCs of constant force ensembles\cite{liuf}, x-coordinate is set to be 
force $f$.}
\label{homopolymerefcs01}
\end{figure}

We investigate how temperature $T$ and monomer number $(N+1)$ affect 
FECs of homogeneous chains of hairpin and secondary structure conformations, 
respectively; see Figs.~\ref{homopolymerefcs01}. Here only positive 
extensions are given. There are common features in these FECs. First, forces $f$ always monotonously 
increases as extensions $x$ increase except the start part. Second, 
the FECs shapes in monotonous regions of any type of conformations are 
very similar at different $N$ value, i.e., $f$ may be the function of 
$x/N$, or relative extension $\rho=x/N$. Finally, at the same extension 
interval, the force changes are faster at higher temperature. To 
check the relationship of force and relative extension $\rho$, FECs of 
different $N$ values  at the same temperature are plotted together in 
terms of $f$ versus $\rho$. We find that these FECs quickly tend to 
asymptotic curves as $N$ increases; these curves are related only with 
temperature and chain conformation type (results are not shown in this 
paper). Differences in FECs between two classes conformations are also 
apparent: FECs of hairpin cases are identical over a longer region of 
extension than those of secondary structure conformations; when 
extensions increase, forces increase continously and smoothly in 
secondary structure cases, while forces in hairpin cases are almost 
invariant, until dramatical jumps happened as extensions reaching their full 
length. 

Comparing FECs with EFCs of homogeneous double-stranded chain molecules 
is valuable. The latter have been computed in Ref.~\cite{liuf} (see 
Fig. 8 therein). It is about equivalence of the two conjugated ensembles: 
the constant  extension and the constant force\cite{lubensky,titantah}. 
We find the FECs and EFCs of two ensembles are almost consistent in the 
monotonous regions, though small differences present. The deviation could 
be expected due to finite $N$-value\cite{titantah}.    

\begin{figure}[htpb]
\begin{tabular}{cc}
\includegraphics[width=0.4\columnwidth]{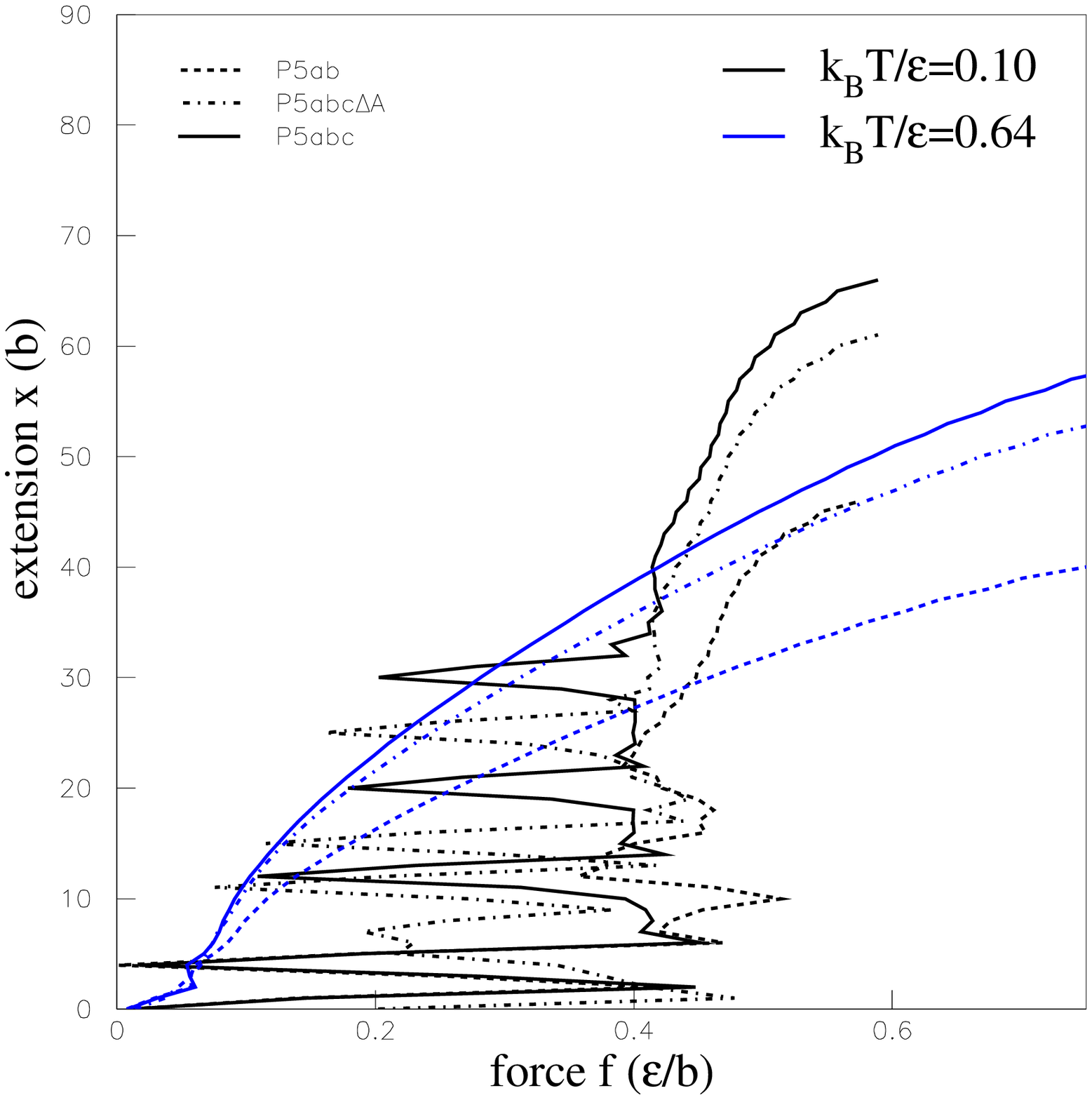}&
\includegraphics[width=0.4\columnwidth]{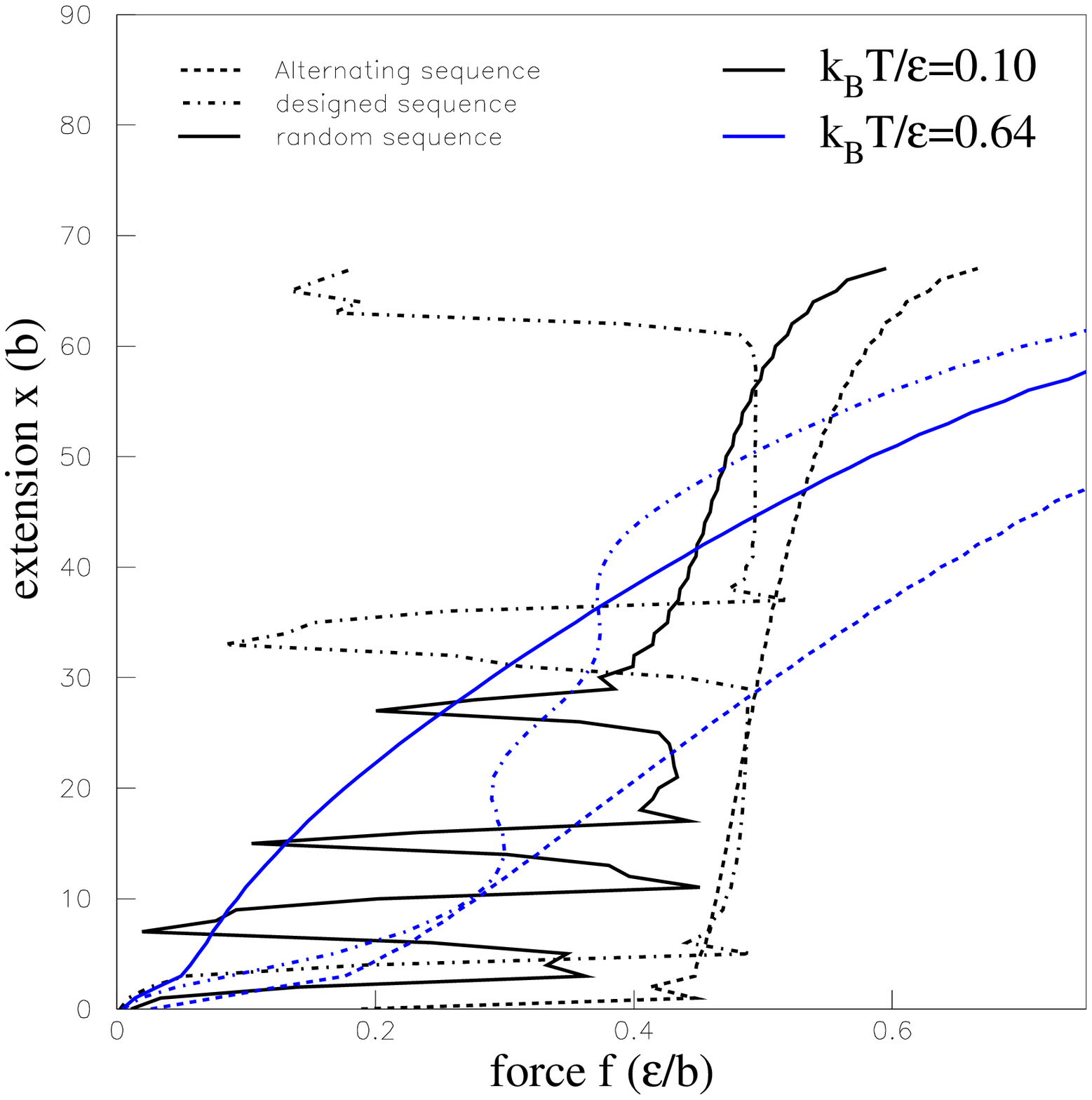}\\
(a)&(b)\\
\end{tabular}
\caption{The FECs of heterogeneous secondary structure chains: 
(a) the sequences are 49-mer P5ab, 65-mer P5abc$\Delta$A and 69-mer P5abc. 
(b) 70-mer sequences are random, alternating and designed. Two colors 
represent FECs at different temperature. }
\label{heterefcs}
\end{figure}

To illustrate effects of specific monomer sequence, FECs of RNA sequences P5ab, 
P5abc$\Delta$A, and P5abc are computed; 
see Fig.~\ref{heterefcs}(a). These sequences come from force unfolding RNA 
experiments studied by Liphardt {\it{et al.}} recently\cite{liphardt}. To 
be very different from FECs of homogeneous chains, the FECs of heterogeneous 
chains are no longer increase monotonously: sawtooth-like oscillations 
present in all sequences. The force behaviors are very similar with 
the force curves observed in unzipping dsDNA experiments\cite{bockelmann}. 
The complex features disappear at higher temperature, such as at 
$0.64\varepsilon/k_B$. We also compare the FECs with the EFCs of force 
stretched the same RNA sequence\cite{liuf}: no oscillations was observed in 
constant force ensembles. Unlike the homogeneous chains, the equivalence of 
two conjugated ensembles is not guaranteed as consideration sequence 
effects. In particular, the force non-monotonicity even survives  
in thermodynamic limit for each sequence realization since the absence of 
self-averaging in biomolecules\cite{bhat}. 

We also calculate FECs of random, alternating and designed sequences; see 
Fig.~\ref{heterefcs}(b). The designed is composed of four elements, A, U, C 
and G arranged by $A\cdots ACCCCU\cdots UC\cdots CAAAAG\cdots G$, where the 
dots represent 15-mer A, U, C and G consecutively. Comparing the FECs of 
the artificial sequences with curves got before, we find the force curves of 
the random sequence is similar with the curves of biomolecular sequences in 
Fig.~\ref{heterefcs}(a); the forces of the alternating sequence are almost 
the same with the results of homogeneous secondary structure chain; 
while the FEC of the designed sequence presents large force reduction at 
extension about a half of full length. From designed sequence arrangement, if  
no extension constraint, the molecule tends to form two independent identical 
hairpins simultaneously. The larger force reduction reflects completely 
collapsing of one hairpin. Interestingly, when force stretching the 
designed sequence, only one jump exhibits since higher cooperativity between 
two hairpins\cite{liuf}.   

\subsection{Contact distribution and asymmetric function}
In order to gain insights into the molecular structure at any extension,  
we compute contact distribution $p(i,j;T,m)$ for every possible 
contact pair $(i,j)$, where $m$ is the discrete value of extension. 
$p(i,j;T,m)$ is determined by the ratio of the conditional partition function 
${\mathcal Z}_N(i,j;T,m)$ for all the conformations that contain contact 
$(i,j)$ and full partition function ${\mathcal Z}_N(T;m)$, i.e.,   
$p(i,j;T,m)={\mathcal Z}_N(i,j;T,m)/{\mathcal Z}_N(T;m)$. As an example, 
the density diagrams of a 70-mer homogeneous hairpin chain at different 
extensions are shown in Fig.~\ref{density}. 

\begin{figure}[htpb]
\begin{tabular}{cc}
\includegraphics[width=0.4\columnwidth]{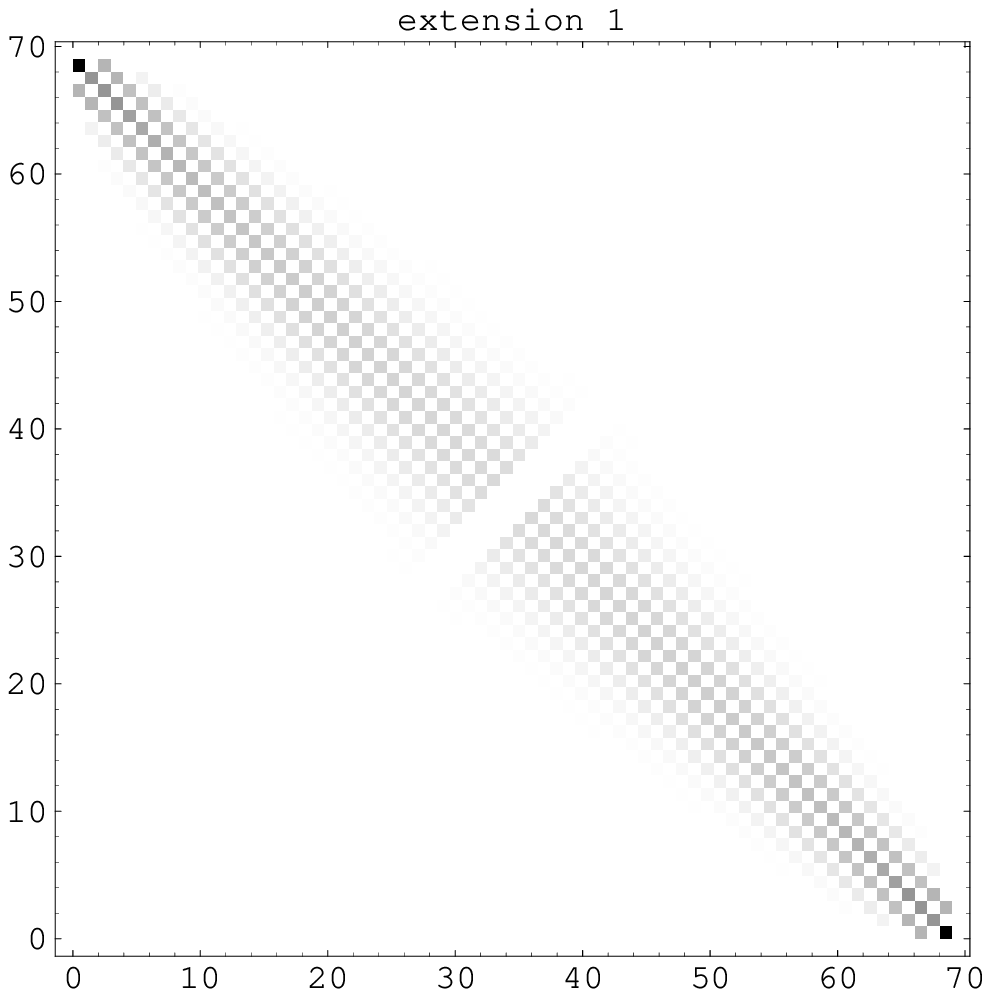}&
\includegraphics[width=0.4\columnwidth]{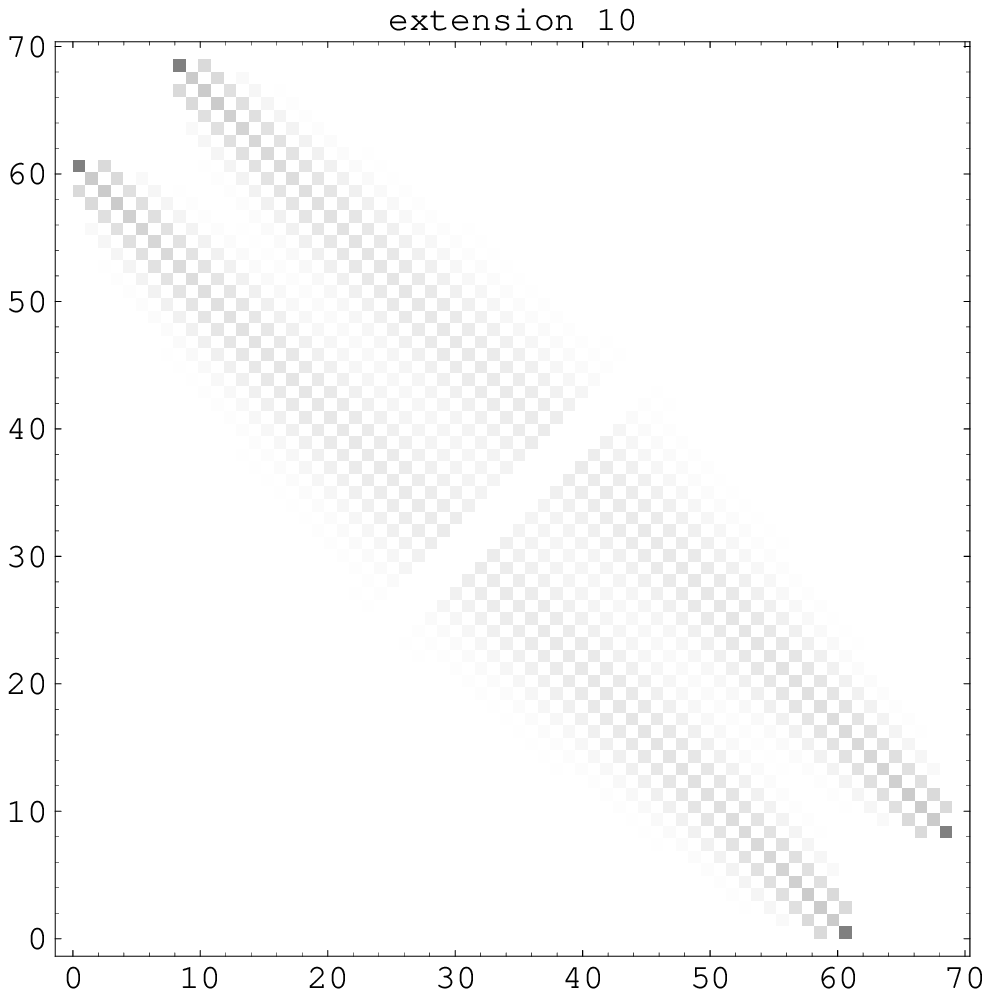}\\
(a)&(b)\\
\includegraphics[width=0.4\columnwidth]{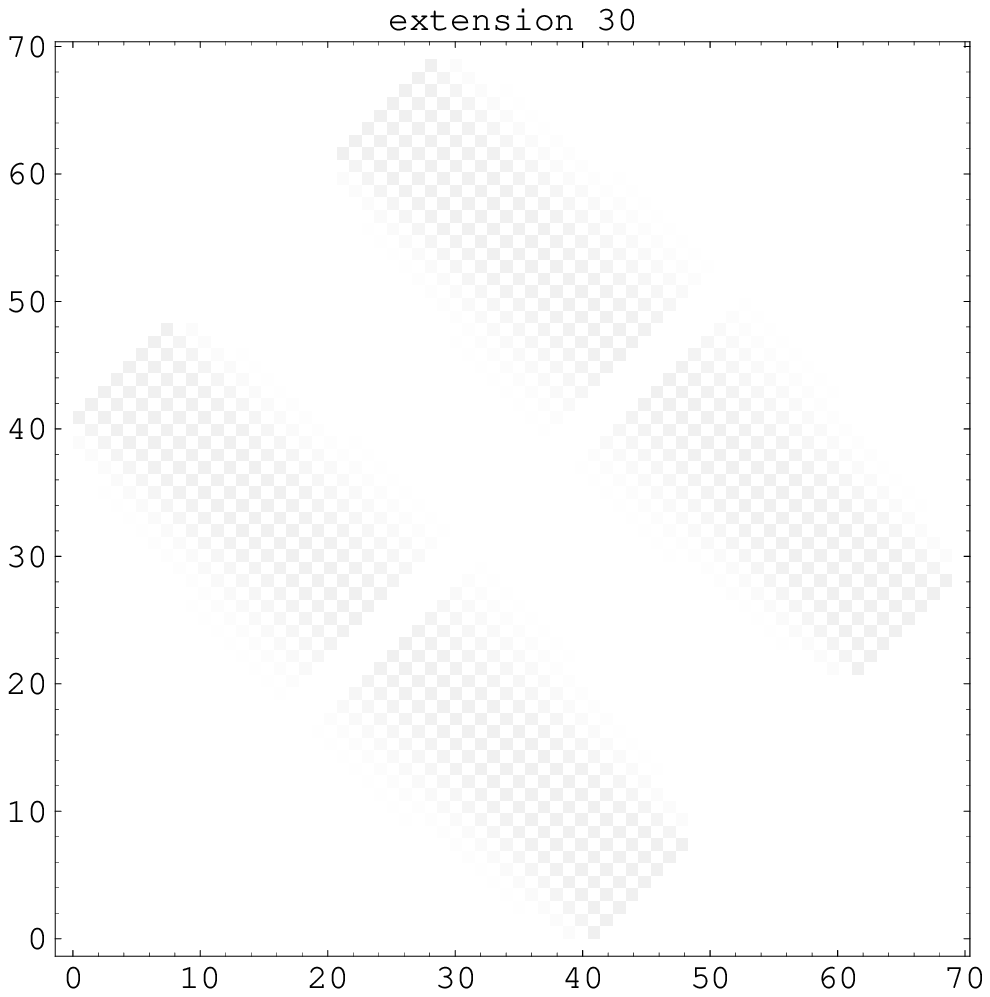}&
\includegraphics[width=0.4\columnwidth]{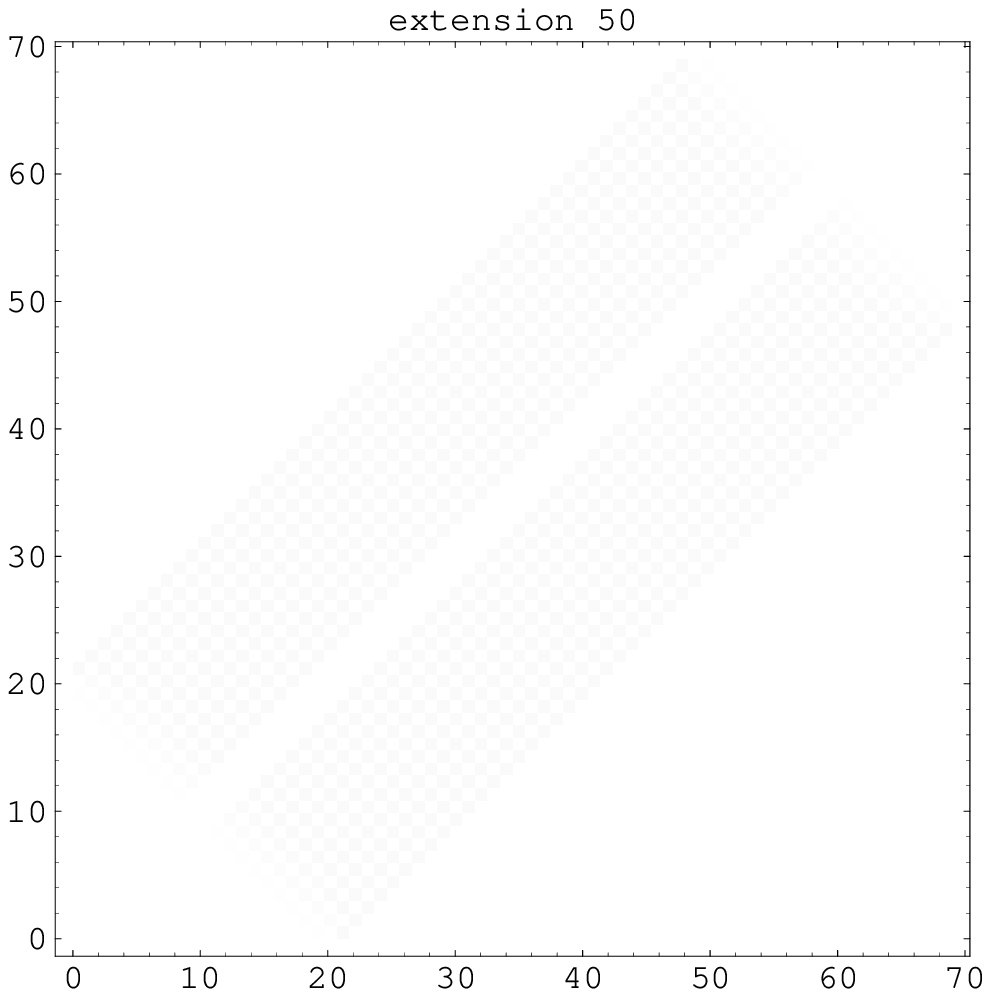}\\
(c)&(d)
\end{tabular}
\caption{The density diagrams for contact probability $p(i,j;T,m)$ of a 70-mer 
homogeneous chain of hairpin conformations at different extensions 1,10,30 
and 50$b$. Here temperature is $0.1 \varepsilon/k_B$. When extensions are 
smaller, such as in (b) and (c), nonzero contacts populate two 
sides of, instead of along the catercorner of the diagrams.}
\label{density}
\end{figure}

\begin{figure}[htpb]
\begin{center}
\includegraphics[width=0.5\columnwidth]{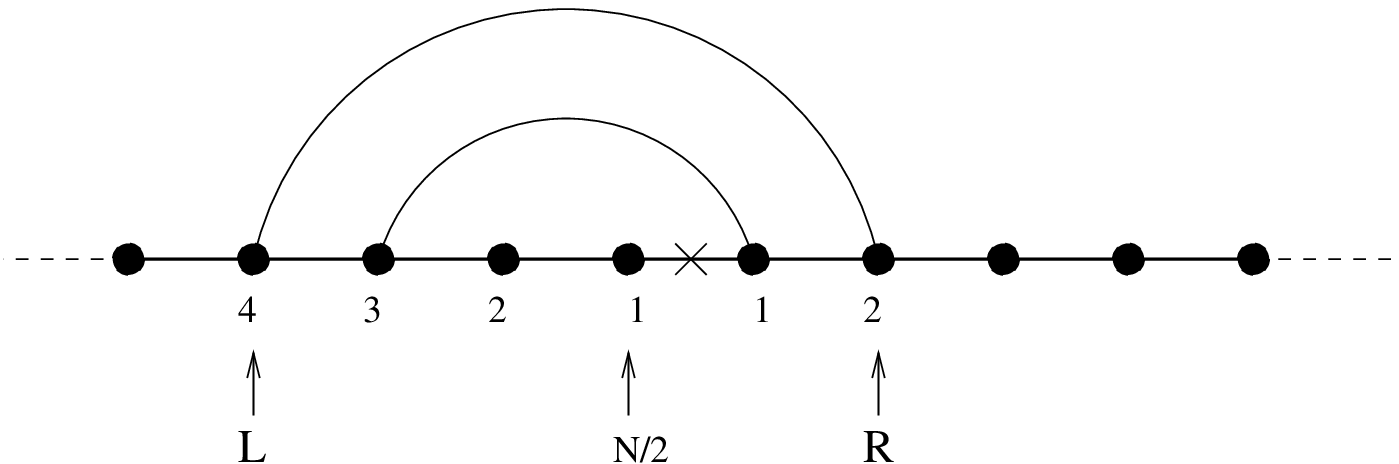}
\caption{Sketch of the AF $S$ definition. $L$ and $R$ are the 
number of monomers apart from the middle position (`x' symbol) of the 
chain to the outmost link of CG region. Here even value of $N$ is 
considered for simplicity. } 
\label{LRgraph}
\end{center}
\end{figure}

It is unexpected that nonzero contacts populate symmetrically two sides of 
the catercorner of diagrams when extensions are smaller, such as 
in Figs.~\ref{density}(b) and (c). It means that the CG regions in 
hairpin conformations tend to locate at two sides of the chain. We did not 
observe the phenomena for the same chain in constant force 
ensembles before. To characterize the phenomena quantitatively, we 
introduce AF $S(L,R)$, 
\begin{eqnarray}
 S(L,R)=\theta(L-R)(1-\frac{min(L,R)}{max(L,R)}),
\label{afdef}
\end{eqnarray}
where $L$ and $R$ are the number of monomers apart from the 
middle position to the outmost link of CG region; see Fig.~ \ref{LRgraph}, 
$\theta$ is SIGN function, and functions min(L,R) and max(L,R) equal the 
minimum and the maximum value of $L$ and $R$, respectively. $S$ values of the 
CGs located completely on the left (right) of middle position are assumed to 
be $+1$ ($-1$), while in conformations containing no contacts, take $S=0$.  
\label{density}

\begin{figure}[htpb]
\begin{tabular}{cc}
\includegraphics[width=0.4\columnwidth]{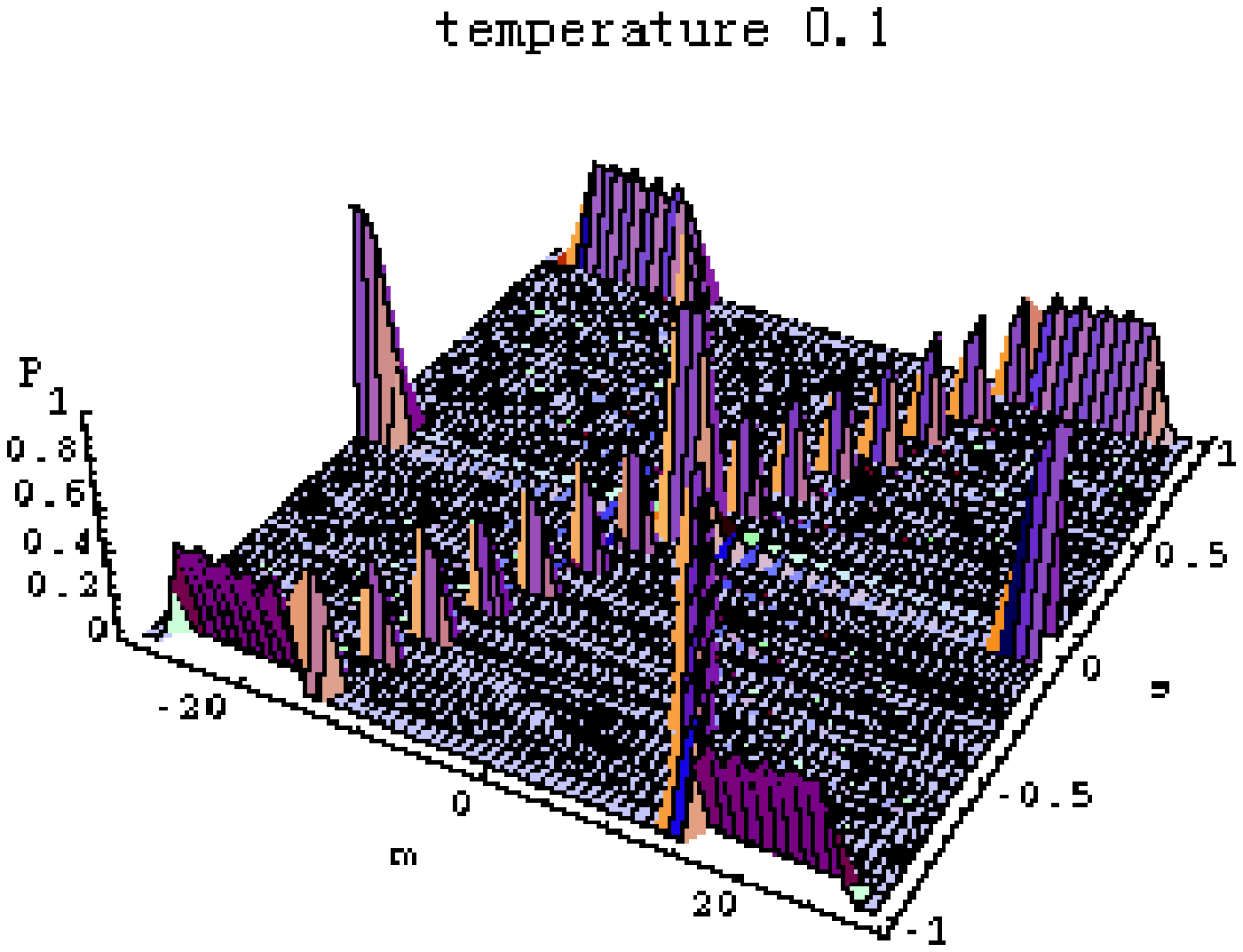}&
\includegraphics[width=0.4\columnwidth]{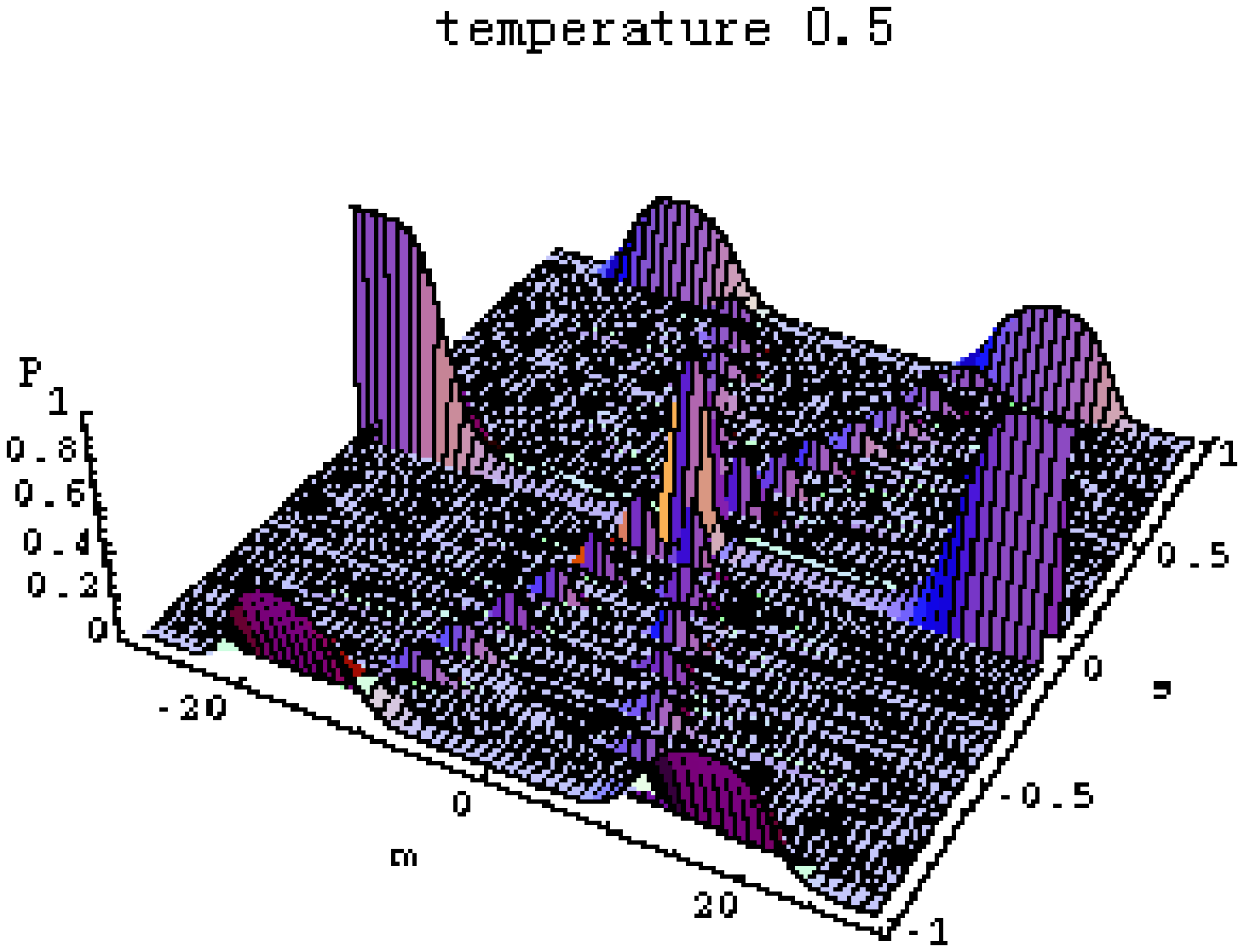}\\
(a)&(b)
\end{tabular}
\caption{The population probabilities $p(m,s;T)$ for 30-mer homogeneous chain 
of hairpin conformations. Temperatures are $0.1$ and $0.5\varepsilon/k_B$. 
Apparent `X' patterns can be observed in figures, which shows that CG 
regions of dominant conformations at these extensions tend to form 
at two sides of the chain. }
\label{asymmetry}
\end{figure}

Then define population probability $p(m,s;T)$ as 
\begin{eqnarray}
p(m,s;T)={\mathcal Z}_N(s;T,m)/{\mathcal Z}_N(T;m),
\end{eqnarray}
where ${\mathcal Z}_N(s;T,m)$ also is the conditional partition function for 
all the conformations whose AF values 
are $s$. Figs.~\ref{asymmetry}(a) and (b) are the distributions 
of 30-mer homogeneous hairpin chain at temperatures $0.1$ and $0.5\varepsilon/k_B$. 
We see that the $p(m,s;T)$ can be divided into four parts, e.g., in the 
Fig.~\ref{asymmetry}(a): at very small extensions (about smaller than 
2$b$), the maximum values of the distribution are at $s=0$, i.e, the CGs 
of the dominant conformations locate on the middle of the chain according 
to the definition of AF; when extensions are larger but still smaller some  
value (about 15$b$), the maximum distribution locate at $0<\pm s<1$. It 
means that CG regions stand two side of the chain. Interestingly, 
these $s$ values stand on two lines (`X' patterns in figures); if 
increasing extensions further (smaller than 26$b$), the $s$ values of 
the maximum distributions are $\pm1$; finally, distributions 
will shift to $s=0$ as extensions reach the full length.            

\begin{figure}[htpb]
\begin{tabular}{cc}
\includegraphics[width=0.4\columnwidth]{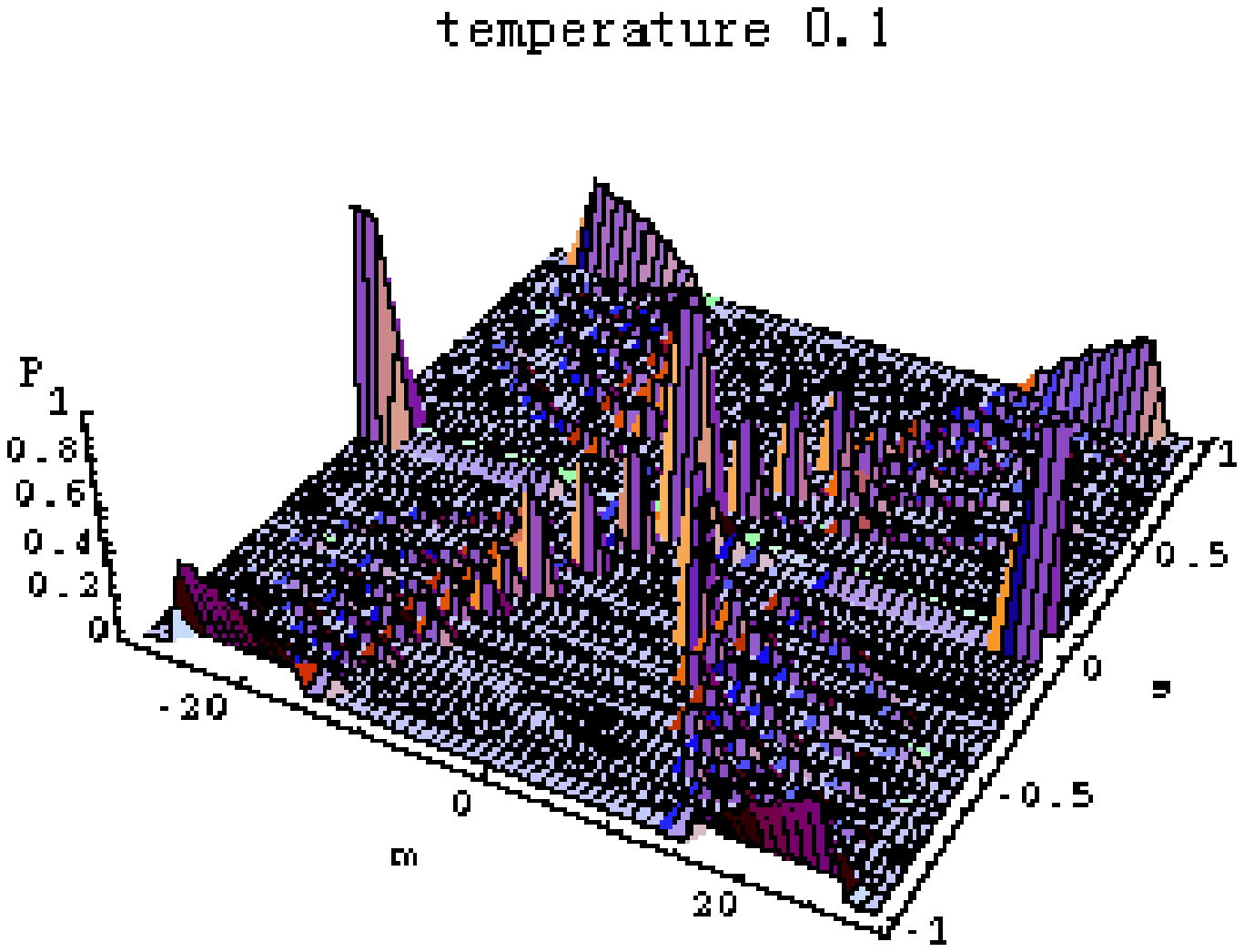}&
\includegraphics[width=0.4\columnwidth]{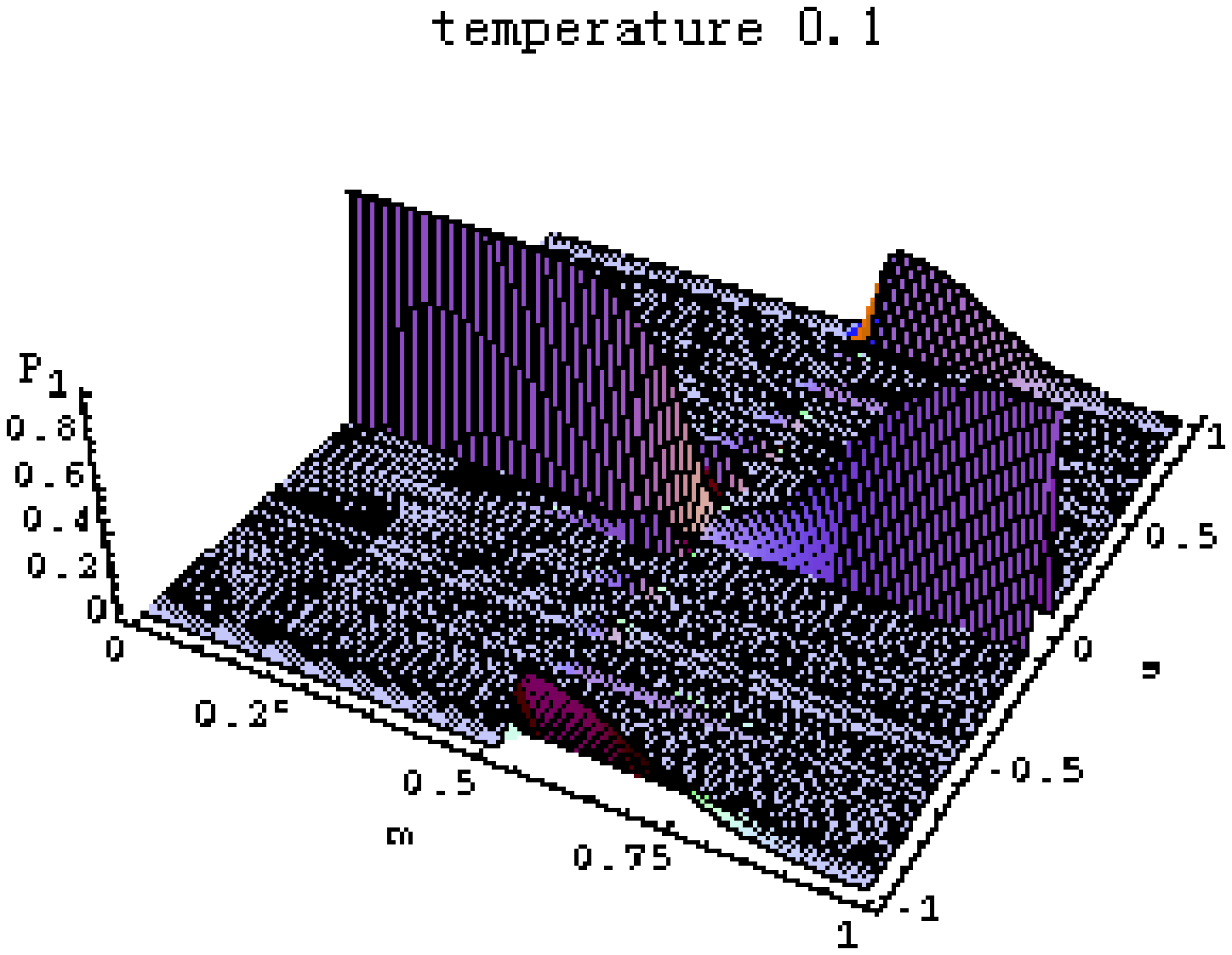}\\
(a)&(b)
\end{tabular}
\caption{(a) The population probabilities $p(m,s;T)$ for 30-mer homogeneous 
chain, where EV interactions between CG regions and two tails 
are considered partially. the `X' in Fig.~\ref{asymmetry} transforms  
into ``butterfly". (b) The population probabilities $p(f,s;T)$ 
at temperature $0.1 \varepsilon/k_B$, where forces are in unit 
$\varepsilon/b$. The extent of $s=\pm1$ becomes narrow as value of $N$ 
increases, since EFCs of the hairpin conformations have first-like behavior.}
\label{psymmetry}
\end{figure}

The behaviors of the distribution at the first and third parts can be 
understandable. 
As extensions are very small or even vanish, the favorable conformations 
of the hairpin are those of most contacts. At lower 
temperatures, say $0.1\varepsilon/k_B$, for homogeneous chain, such 
conformations are forming consecutive contacts: $(0,N),(1,N-1),\cdots,
(N/2-1,N/2+2)$. Hence, the AF value of the maximum distribution is $0$. While 
extensions is larger enough, or temperatures are higher, if the number of 
monomers of the CG is smaller than $N/2$, the total number of conformations 
whose CGs present in two sides must be far larger than that whose CGs 
stand in the middle of the chain. According to AF definition, the 
distributions naturally concentrated on $s=\pm1$; this is only combination 
result. But why in a range of extensions, the dominant conformations 
tend to form at two sides of the chain, or AF is neither 0 nor $\pm1$, even 
if their CGs across over the middle position? Considering that contact 
interactions along the chain is uniform, energy must not be the reason. 
We believe that the nonuniform EV interactions of the CG regions and two 
tails with different lengths (see Fig.~\ref{hairpinextn}) lead to the tendency: 
under fixed extension and the same number of CG size, the number of 
conformations having two tails with lengths $l_1$- and $l_2$-mer is smaller 
than the number of conformations having one $(l_1+l_2)$-mer tail; or given the 
same number of tail monomers, the EV effects of CG region are more strong with 
two tails than with one tail. At temperature $0.1\varepsilon/k_B$,  
given the extension $x$, because of the most number of contacts encouraged, 
then the maximum CG size $(N-x)$-mer is favorable. If supposing one tail 
existing at left, one easily give the following relation: 
$s\approx\pm2x/N=\pm2 \rho$. Comparing it with Fig.~\ref{asymmetry}(a), the 
formula agree with the pattern very well: the slopes of two lines are about $\pm2$. 
On the other hand, Fig.~\ref{asymmetry}(b) shows that the slopes increase with  
temperature. In fact, at higher temperature $T_h$, the maximum CG size is unfavorable 
for thermal fluctuation. For the same extension, given that $L$ and $R$ values at 
origin temperature are $l$ and $r$ respectively. Supposing $t$ contacts broken at 
$T_h$, then the s-value changes from $1-r/l$ to $1-(r-t)/(l-t)$. Apparently, AF is 
the increasing function of $t$, while $t$ also be an increasing function of 
temperature.        
As a comparison, we partially cancel EV interactions: when one length 
of two tails is larger than some value, here $5b$ is taken, the interactions 
will be screened. The distribution of 30-mer chain is computed again 
under this the requirement; see Fig.~\ref{psymmetry}(a). The maximum values of 
distribution at a fixed extension are not located on a unique $s$, or 
CGs distribute on the chain uniformly. Because the lines with slopes 
$\pm2$ represent the maximum asymmetry at temperature $0.1\varepsilon/k_B$, other 
AF values must stand between them (see ``butterfly" pattern in 
Fig.~\ref{psymmetry}(a)).

It may provide useful insights into the force stretching problem using asymmetric 
concept. Define $p(f,s;T)= 
{\cal Q}_N(f,s;T)/{\cal Q}_N(T;f)$, where conditional partition function 
${\cal Q}_N(f,s;T)$ is 
for all conformations whose asymmetric factor is $s$, and ${\cal Q}_N(T,F)$ 
is full partition function\cite{liuf}. Fig.~\ref{psymmetry}(b) is the diagram 
of $p(f,s;0.1)$ of 30-mer homogeneous hairpin chain. The $p(f,s;T)$ is 
very different from $p(x,s;T)$, e.g., the AF values of the former 
change from $0$ to $\pm1$, and reenter into $0$ in narrow force ranges, whereas 
for the latter, the changes of the AF are continuous and slowly.  
At temperature $0.1\varepsilon/k_B$, when forces are smaller than 
$0.48\varepsilon/b$, states having the most number of contacts are favorable, 
which means most possibility values of AF are $0$. Because of higher cooperativity 
in hairpin conformations, the structures are collapsed in small force range as 
$f>0.48\varepsilon/b$\cite{liuf}. It means AF vanishes again. The temporary 
$s=\pm1$ should be the results of finite chain length.    
  
\subsection{Force flipping phenomena}
\begin{figure}[htpb]
\begin{tabular}{cc}
\includegraphics[width=0.4\columnwidth]{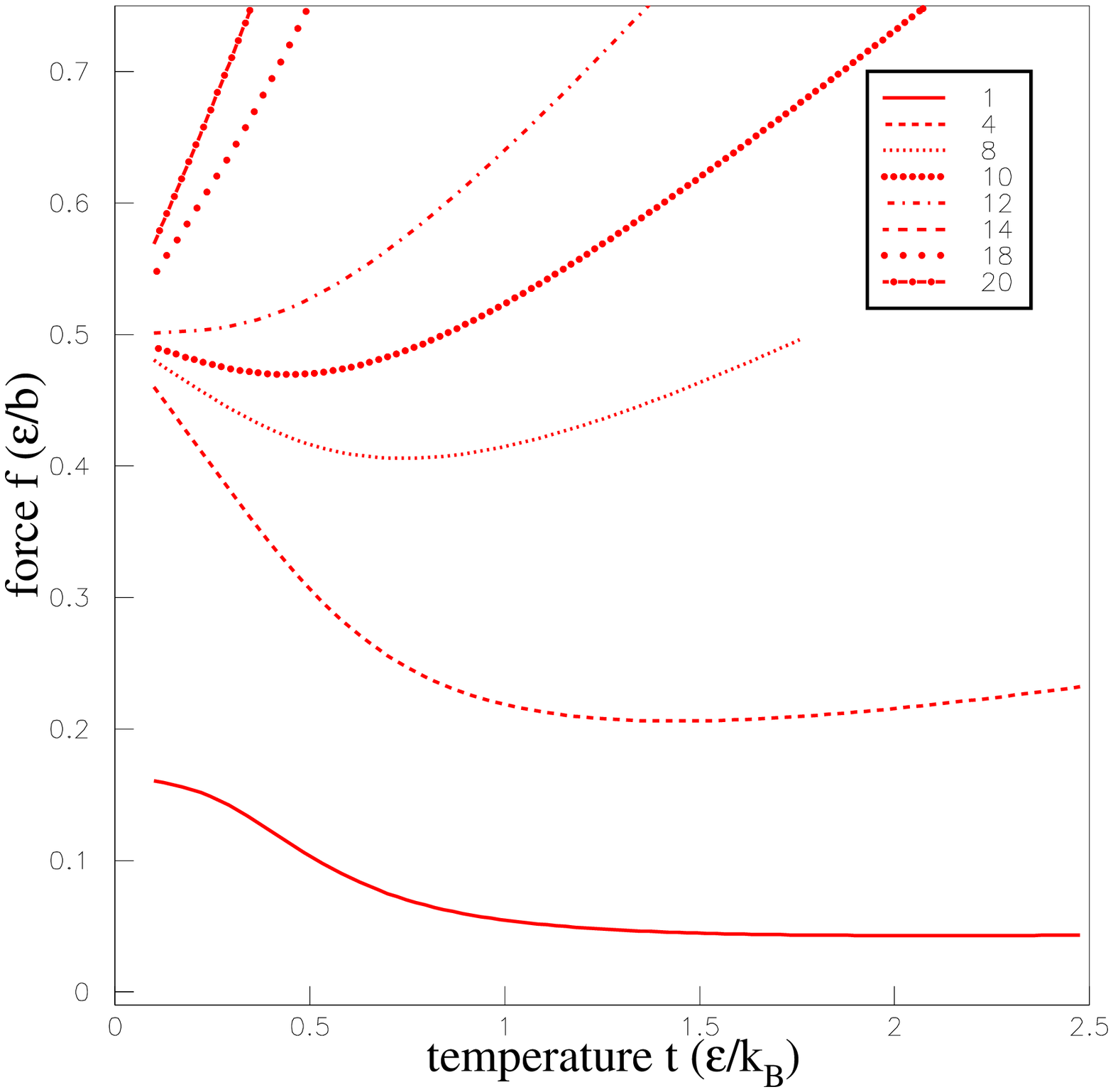}&
\includegraphics[width=0.4\columnwidth]{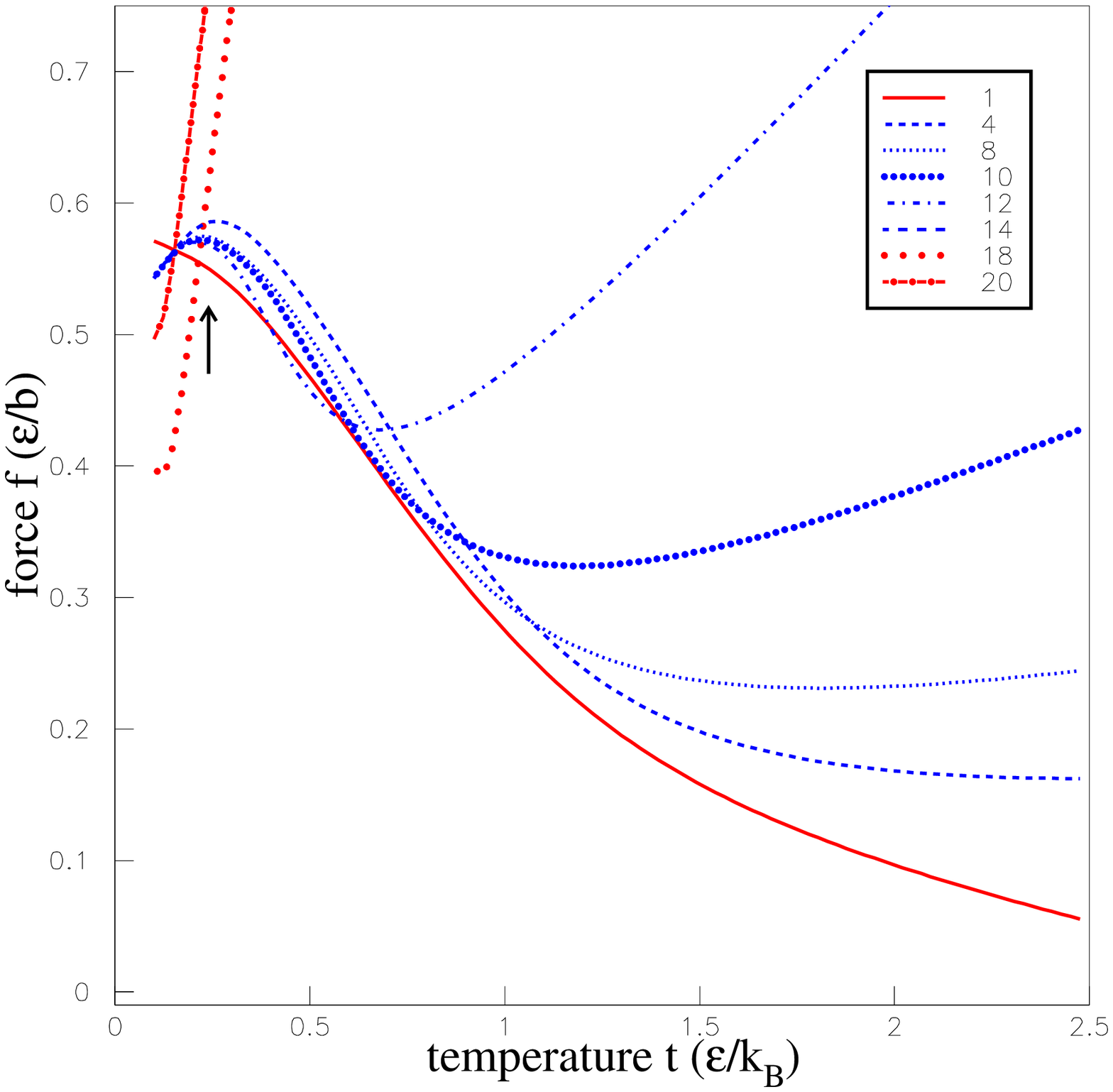}\\
(a)&(b)
\end{tabular}
\caption{The TFCs of 25-mer chains of secondary structure (a) and hairpin 
conformations (b) at different fixed extensions.  
where the sequences are homogeneous. Arrow in (b) points out the convex, which 
demonstrates the appearance of force-flipping transition. }
\label{reenter}
\end{figure}

Finally, we explore the force dependence on temperatures as fixed extension. 
The studies are related with the re-entering transition studied in 
unzipping dsDNA\cite{marenduzzo,orlandini}. The transition means that if 
one fixes external force at a value in a finite range and decreases 
temperature, beginning with stretched state, dsDNA will first collapse to a 
globular state; while temperature is lowered further, dsDNA will re-enter 
stretched state again. We have discussed the phenomenon in constant force 
stretched homogeneous chains of hairpin and secondary structure 
conformations in Ref.~\cite{liuf}: the re-entering transition only presents 
in hairpin case. To our knowledge, the conjugated phenomena of re-entering 
transitions in constant extension ensembles are not investigated before. 
Our model can be used to check the problem by numerical computing FTCs of 
homogeneous chains of hairpin and secondary structure conformations, 
respectively. The results of 25-mer chains are shown in 
Fig.~\ref{reenter}: force flipping (the arrow therein) presents at lower 
temperature in hairpin conformations, but is not observed in secondary 
structure conformations. We believe that the flipping just corresponds to 
re-entering transitions in constant force ensembles. This results could be  
explained qualitatively. First, the temperatures of flipping and re-entering 
transitions are almost the same, i.e., about $0.23\varepsilon/k_B$\cite{liuf}. 
Second, like the absence of flipping in secondary structures, re-entering 
transitions of this conformations are not observed in constant force ensembles. 
Finally, because of the equivalence of constant extension and constant force 
ensembles of homogeneous chains, the FTCs at fixed extensions could be 
derived from the extension-temperature curves (ETCs) at fixed force. In the 
fixed force ensembles, if there is a dip at lower temperature $T_r$ in the ETC 
of a given force, then at least two temperatures located at two sides 
of $T_r$ correspond to the same extensions $x_o$. If the dip rises as the force 
increasing, the extension $x_o$ will have a serious of temperatures standing 
in two sides $T_r$, which will form a convex in FTC at $T_r$. The dips in ETCs 
of hairpin conformations do have such characteristics\cite{liuf}. We can use the 
same approach to understand FTCs behaviors of secondary structure conformations. 

\section{Summary}
\label{conclusion}
In this paper, we propose a statistical model of constant extension 
ensembles of double-stranded chain molecules. Unlike several theoretical 
models proposed previously, specific monomers sequence and 
EV interaction are exactly included. Using the model, we 
investigate how FECs depend on chain length, sequence, and 
structure. In addition, the model can relate FECs with molecule structure 
directly. The structures of hairpin conformations reveal that CG regions 
tend to form at two sides of chains as the extensions are in some range.  
This unexpected phenomena have never been reported before. Through 
introducing AF and analyzing in detail, we contribute the phenomena to  
EV interactions between CG regions and tails. We also explore the 
conjugated transitions of re-entering in constant extension ensembles, 
the force-flipping phenomena, which only present in hairpin conformations. 
Because our aim in this paper is to illustrate, the model is largely 
simplified, e.g., the 
chain is restricted on 2D lattice, chain stiffness (bending) and elasticity 
are neglected, and no stacking interaction is involved. However, our results 
still give many important physical information. especially the importance 
of the long-range EV interaction. In future work, we will 
replace the projection extension by real EEDs. It is interesting to 
see whether the position tendency of CGs is still observed.  

\begin{acknowledgments}
We are grateful to Dr. Shi-jie Chen, and Prof. H.-W. Peng for much useful  
discussion.
\end{acknowledgments}

\appendix
\section{calculating extensions of the hairpin conformations}
\label{endtoendhairpin}
\begin{figure}[htpb]
\begin{center}
\includegraphics[width=1.\columnwidth]{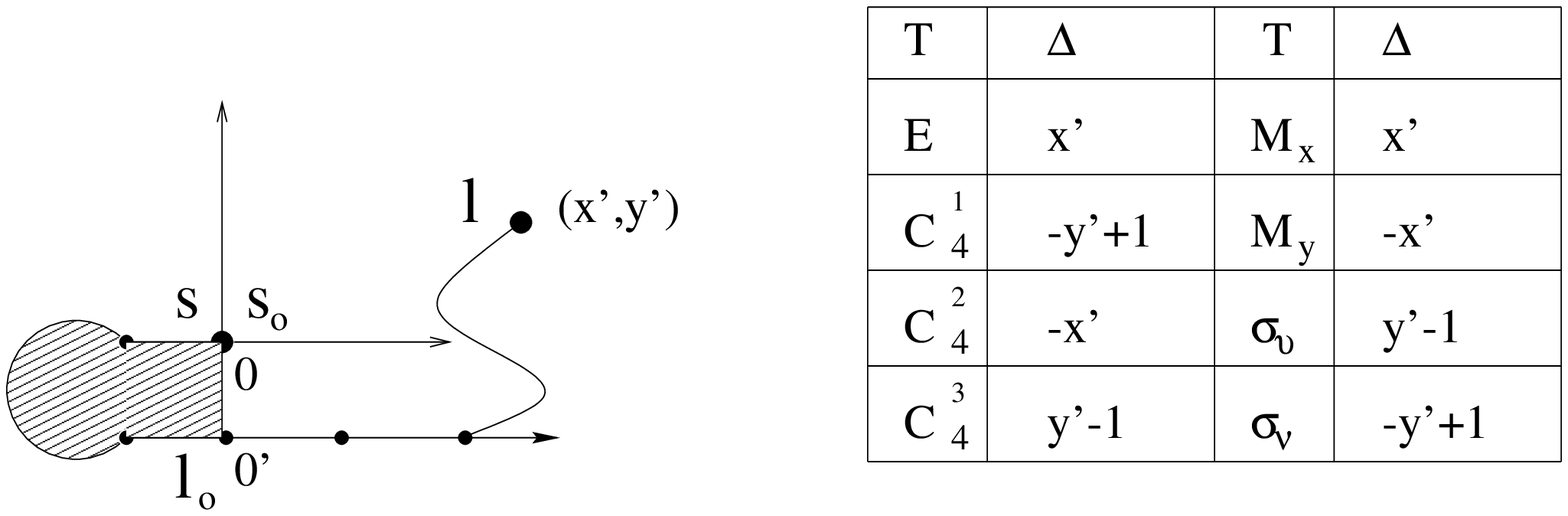}
\caption{One class of conformations for the 2nd tail $(l_o,l)$ and closed 
graph $(s_0,l_o)$ containing outmost type 1 link, where the tail type 
is stiff and extends upward \protect\cite{chen98}. The end $l$ located at 
$(x',y')$ on frame $O'$ is translated to point $(x',y'-1)$ 
on frame $O$, and the corresponding extension $\Delta$ is $x'$. We can  
distribute the conformation to lattice plane by eight square symmetry 
transformations (see Fig.~\ref{hairpinextn}). The extension $\Delta$s are 
tabulated respectively, where the character ``T" represents transformation 
elements. 
}
\label{complextail}
\end{center}
\end{figure}

For hairpin conformations, the conformations of tails are determined by 
two factors: the outermost link type and the length of closed graph. To 
account for EV interactions, the tails have been classified into two 
types, stiff (s) and flex (f)\cite{chen98}. 
Correspondingly, the number of conformations is $n^{t_i}(N_i)$, where $t_i$ 
(s or f) are the type of $i$th (1st or 2nd) tail with length $N_i$, or 
$N_i$-step OSAW. In order to calculate extensions of the whole chain , we introduce 
additional definitions: $n_x^{t_i}(N_i;m)$, the number of conformations 
for $i$th OSAW of type $t_i$ whose final $x$ coordinates are $m$; and 
$n_y^{t_i}(N_i;m)$ is similar except $y$ coordinate. It is ease to find  
$n^{t_i}(N_i)=\sum_m n_x^{t_i}(N_i;m)=\sum_m n_y^{t_i}(N_i;m)$. 
Here we illustrate the calculation of the simplest diagonal matrix element 
${\bf \omega} (l,l_0;s_0,s|\Delta)_{11}$, where the outmost link is of type $1$ 
and $s=s_0$, $l\ne l_0$. Because other more general cases have more cumbersome 
formula, we are not listed them in the present paper. 

There are six classes of possible conformations for the simplest CG-tail 
complexes\cite{chen98}. In Fig.~\ref{complextail}, we show one of them. 
The length of the 2nd tail then is $N_2=l-l_o$. Since the length of CG also 
affects EV interactions, we distinguish different CG sizes. 

(1) $l_o-s_o=3$.
All six classes of conformations are viable. Therefore, 
\begin{eqnarray}
{\bf \omega}(l,l_0;s_0,s|\Delta)_{11}&=&2\left[ n_y^{f_2}(N_2;\Delta) + n_y^{f_2}(N_2;-\Delta) + n_y^{f_2}(N_2;\Delta-1) + n_y^{f_2}(N_2;-(\Delta+1))\right.
\nonumber\\
&&+n_x^{f_2}(N_2;\Delta) + n_x^{f_2}(N_2;-\Delta) + n_x^{f_2}(N_2;\Delta-1) + 
n_x^{f_2}(N_2;-(\Delta+1))\nonumber\\ 
&&+2n_y^{s_2}(N_2;\Delta) + 2n_y^{s_2}(N_2;-\Delta) + n_y^{s_2}(N_2;\Delta-1) 
+ n_y^{s_2}(N_2;-(\Delta+1))\nonumber\\
&&+n_y^{s_2}(N_2,\Delta+1)+ n_y^{s_2}(N_2;-(\Delta-1))\nonumber\\
&&+2\left(n_x^{s_2}(N,\Delta)+n_x^{s_2}(N,-\Delta)+n_x^{s_2}(N,\Delta-1)+n_x^{s_2}(N,-(\Delta+1)) \right)\nonumber\\
&&\left.-\left(2\delta_{0,\Delta}+\delta_{1,\Delta} +\delta_{-1,\Delta}+
\delta_{N_2,\Delta}+\delta_{-N_2,\Delta}+\delta_{N_2+1,\Delta}+\delta_{-(N_2+1),\Delta}\right) \right],
\end{eqnarray}
where coefficient $2$ is {\it degeneracy degree} along projection, 
the negative part is to eliminate the straight tail conformations, which 
are counted repeatedly.

(2) $l_o-s_o>3$.
There are five classes of conformations are involved. We have
\begin{eqnarray}
{\bf \omega}(l,l_0;s_0,s|\Delta)_{11}&=&2\left[ 
n_y^{f_2}(N_2;\Delta) + n_y^{f_2}(N_2;-\Delta) + n_y^{f_2}(N_2;\Delta-1) + 
n_y^{f_2}(N_2;-(\Delta+1))\right.\nonumber\\
&&+n_x^{f_2}(N_2;\Delta) + n_x^{f_2}(N_2;-\Delta) + n_x^{f_2}(N_2;\Delta-1) + 
n_x^{f_2}(N_2;-(\Delta+1))\nonumber\\ 
&&+n_y^{s_2}(N_2;\Delta) + n_y^{s_2}(N_2;-\Delta) + n_y^{s_2}(N_2;\Delta-1) 
+ n_y^{s_2}(N_2;-(\Delta+1))\nonumber\\
&&+n_y^{s_2}(N_2,\Delta+1)+ n_y^{s_2}(N_2;-(\Delta-1))\nonumber\\
&&+2\left(n_x^{s_2}(N,\Delta)+n_x^{s_2}(N,-\Delta)\right)+ n_x^{s_2}(N,\Delta-1)+n_x^{s_2}(N,-(\Delta+1))\nonumber\\
&&\left.-\left(\delta_{1,\Delta} +\delta_{-1,\Delta}+
\delta_{N_2,\Delta}+\delta_{-N_2,\Delta}\right) \right].
\end{eqnarray}


\begin{thebibliography}{99}

\bibitem{smith96}
S. B. Smith, Y. J. Cui, and C. Bustamante, \sci{271}{795}{1996}.

\bibitem{maier}
B. Maier, D. Bensimon, and V. Croquette, \pnas{97}{12002}{2000}.

\bibitem{zlatanova}
J. Zlatanova, S. M. Lindsay, and S. H. Leuba, Prog. Biophs. Mol. Biol.
{\bf 74}, 37 (2000).

\bibitem{riefp}
M. Rief, M. Gautel, F. Oesterhelt, J. M. Fernandez, and H. E. Gaub,
\sci{276}{1109}{1997}.

\bibitem{visscher}
K. Visscher, M. J. Schnitzer, and S. M. Block, \nat{400}{184}{1999}.

\bibitem{bockelmann} 
U. Bockelmann, b. Essevaz-Roulet, V. Viasnoff, and F. Heslot, 
\bio{82}{1537}{2002}.

\bibitem{liphardt} 
J. Liphardt, B. Onoa, S.B. Smith, I.J. Tinoco, and C. Bustamante, 
\sci{292}{733}{2001}.

\bibitem{rief}
M. Rief, H. Clausen-Schaumann, and H. E. Gaub, \nats{6}{346}{1997}.

\bibitem{marko}
J. F. Marko and E. D. Siggia, \mol{28}{8759}{1995}.

\bibitem{zhou1}
H.-J. Zhou, Y. Zhang, and Z.-C. Ou-Yang, \prl{114}{8694}{2001}.

\bibitem{montanari} 
A. Montanari and M. M$\acute{e}$zard, \prl{86}{2178}{2001}.  

\bibitem{lubensky} 
D. K. Lubensky and D. R. Nelson, \pre{65}{031917}{2002}, and references 
therein.

\bibitem{gerland} 
U. Gerland, R. Bundshuh and T. Hwa, \bio{81}{1324}{2001}.

\bibitem{titantah}
J. T. Titantah, \pra{60}{7010}{1999}.

\bibitem{bhat}
S. M. Bhattacharjee and D. Marenduzzo, \jpa{35}{L349}{2002}

\bibitem{chen95} 
S.-J. Chen and K. A. Dill, \jcp{103}{5802}{1995}.

\bibitem{chen98} 
S.-J. Chen and K. A. Dill, \jcp{109}{4602}{1998}.


\bibitem{dessinges}
M.-N. Dessinges, B. Maier, Y. Zhang, M. Peliti, D. Bensimon and
V. Croquette, \prl{89}{248102}{2002}.


\bibitem{marenduzzo}
D. Marenduzzo {\it et al.}, \prl{88}{028102}{2002}.

\bibitem{orlandini}
E. Orlandini {\it et al.}, J. Phys. A {\bf 34}, L751 (2001), 
and references therein.

\bibitem{munoz}
V. Munoz, P. A. Thompson, J. Hofrichter, and W. A. Eaton, \nat{390}{196}{1997}.

\bibitem{liuf}
F, Liu, L.R. Dai and Z.-C. Ou-Yang, (2002), cond-mat/0212268.

\bibitem{fisher66}
M. E. Fisher, \jcp{44}{616}{1966}.

\end{thebibliography}
\end{document}